%% file: main_paper.tex
\documentclass[journal]{IEEEtran}

\usepackage{cite}
\usepackage{hyperref}
\usepackage{amsmath,amssymb,amsfonts}

\usepackage{algorithmic}
\usepackage{textcomp}
\usepackage{amsthm}
\usepackage{balance}
\usepackage{afterpage}
\usepackage{xcolor}
\usepackage[pdftex]{graphicx}
\usepackage[caption=false]{subfig}
\usepackage{amssymb}
\usepackage[hyperfirst=false]{glossaries}
\usepackage{booktabs}
\usepackage[capitalise, noabbrev]{cleveref}
\usepackage{siunitx}
\def\BibTeX{{\rm B\kern-.05em{\sc i\kern-.025em b}\kern-.08em
    T\kern-.1667em\lower.7ex\hbox{E}\kern-.125emX}}

\setkeys{glslink}{hyper=false}
 
\newacronym{nn}{NN}{Neural Network}
\newacronym{cnn}{CNN}{Convolutional Neural Network}
\newacronym{gan}{GAN}{Generative Adversarial Network}
\newacronym{svm}{SVM}{Support Vector Machine}
\newacronym{sar}{SAR}{Synthetic Aperture Radar}
\newacronym{iw}{IW}{Interferometric Wide Swath}
\newacronym{grd}{GRD}{Ground Range Detected}
\newacronym{dbl}{DBL}{Distance Based Logistic}
\newacronym{prnu}{PRNU}{Photo Response Non Uniformity noise}
\newacronym{gmm}{GMM}{Gaussian Mixture Model}
\newacronym{be}{BE}{Baseline Extractor}
\newacronym{sae}{SAE}{SAR Adapted Extractor}
\newacronym{asae}{ASAE}{Augmented SAR Adapted Extractor}
\newacronym{dncnn}{DnCNN}{Denoising Convolutional Neural Network}
\newacronym{em}{EM}{Expectation Maximization}
\newacronym{fed}{FED}{Fingerprint Extraction Dataset}
\newacronym{set1}{SD1}{Spliced Dataset 1}
\newacronym{set2}{SD2}{Spliced Dataset 2}
\newacronym{iou}{IOU}{Intersection Over Union}
\newacronym{tp}{TP}{True Positives}
\newacronym{fp}{FP}{False Positives}
\newacronym{tn}{TN}{True Negatives}
\newacronym{fn}{FN}{False Negatives}
\newacronym{tpr}{TPR}{True Positive Rate}
\newacronym{fpr}{FPR}{False Positive Rate}

\theoremstyle{definition}

\newcommand{\unet}{U-Net}

\begin{document}
\title{Amplitude SAR Imagery Splicing Localization}
\author{{Edoardo Daniele Cannas},
{Nicolò Bonettini},
{Sara Mandelli},
{Paolo Bestagini} and
{Stefano Tubaro}
\thanks{E.D. Cannas, N. Bonettini, S. Mandelli, P. Bestagini and S.Tubaro are with the Dipartimento di Elettronica, Informazione e Bioingegneria, Politecnico di Milano, Milano 20133, Italy, e-mail: name.surname@polimi.it.}}
\null
\newpage

\begin{figure*}
\centering
\includegraphics[width=2\columnwidth]{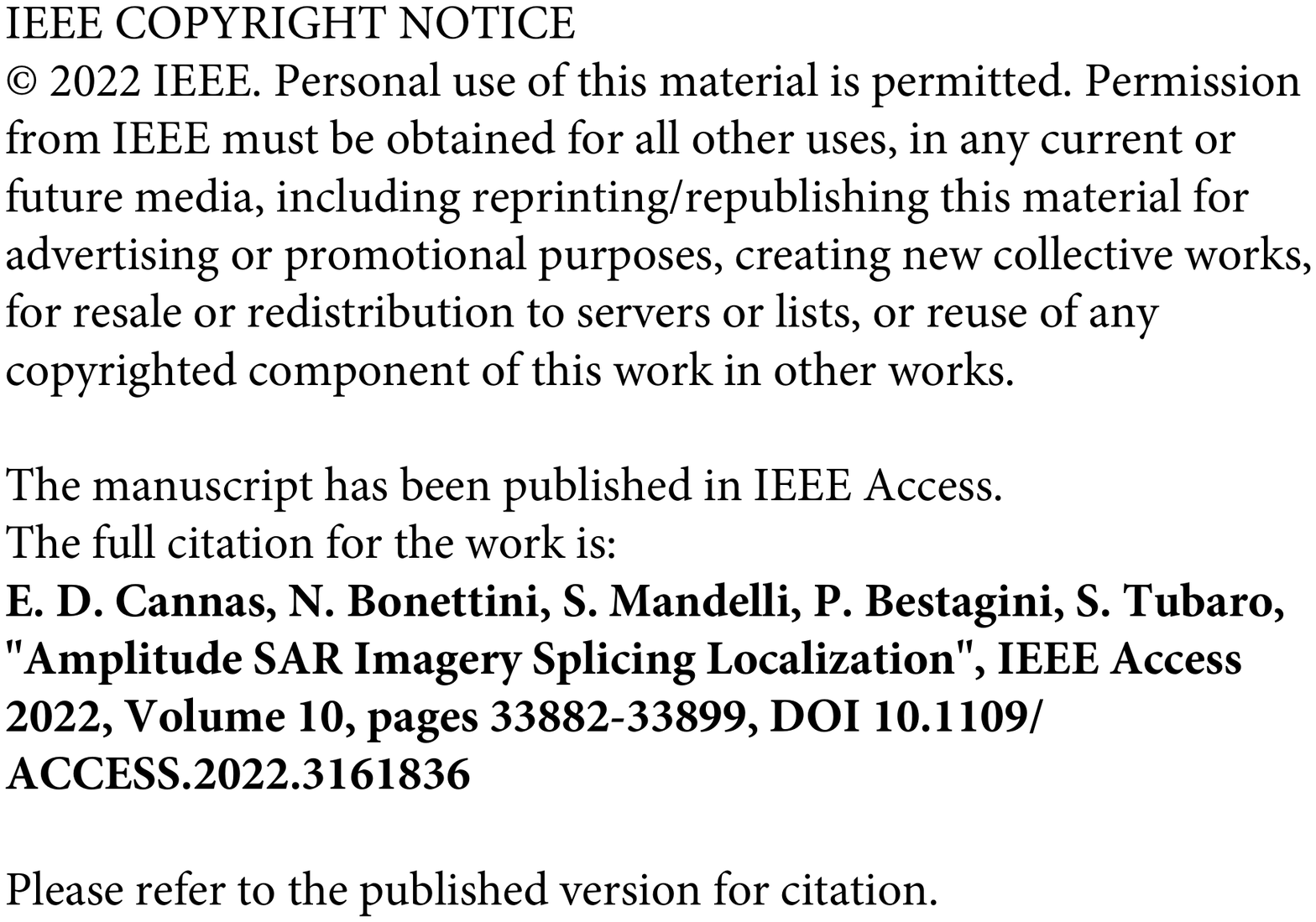}
\end{figure*}
\newpage

\maketitle
\begin{abstract}
Synthetic Aperture Radar (SAR) images are a valuable asset for a wide variety of tasks.
In the last few years, many websites have been offering them for free in the form of easy to manage products, favoring their widespread diffusion and research work in the SAR field.
The drawback of these opportunities is that such images might be exposed to forgeries and manipulations by malicious users, raising new concerns about their integrity and trustworthiness.
Up to now, the multimedia forensics literature has proposed various techniques to localize manipulations in natural photographs, but the integrity assessment of SAR images was never investigated.
This task poses new challenges, since SAR images are generated with a processing chain completely different from that of natural photographs. This implies that many forensics methods developed for natural images are not guaranteed to succeed. 
In this paper, we investigate the problem of amplitude SAR imagery splicing localization.
Our goal is to localize regions of an amplitude SAR image that have been copied and pasted from another image, possibly undergoing some kind of editing in the process. 
To do so, we leverage a Convolutional Neural Network (CNN) to extract a fingerprint highlighting inconsistencies in the processing traces of the analyzed input. Then, we examine this fingerprint to produce a binary tampering mask indicating the pixel region under splicing attack. Results show that our proposed method, tailored to the nature of SAR signals, provides better performances than state-of-the-art forensic tools developed for natural images.
\end{abstract}


\input{sec_introduction}
\input{sec_background}

\input{sec_problem}

\input{sec_method}
\input{sec_setup}
\input{sec_results}
\input{sec_conclusions}


\ifCLASSOPTIONcaptionsoff
  \newpage
\fi

\bibliographystyle{IEEEtran}
\bibliography{biblio}

\end{document}

%% file: sec_introduction.tex
\section{Introduction}
\label{sec:intro}

Due to the lively development of Internet-based communication systems, the diffusion and sharing of multimedia content (i.e., digital images, videos or audio clips) have become part of our daily life.
At the same time, we have become extremely acquainted in using tools for editing these objects.
Doubts regarding whether the content we are enjoying is genuine or not are frequent every day.
Indeed, from politics \cite{Agarwal_2019_CVPR_Workshops} to everyday life experience \cite{recognizeFakeMedia}, the areas where fake media could possibly harm are many.


In this vein, multimedia forensics researchers aim at developing techniques to retrieve information about the multimedia object at hand.
For instance, they are interested in verifying the integrity and trustworthiness of data, spotting manipulated multimedia content and localizing possible forgeries.
Forensics researches typically tackle these problems by considering a simple principle: during the data life-cycle, various non-invertible operations are performed. Each operation leaves a peculiar trace, or footprint, that can be exploited to expose and localize a specific editing. 

The main efforts have been historically directed towards the analysis of digital images \cite{Stamm2013}.
Many techniques have been developed to detect traces left by specific operations executed on the whole picture. For instance, the authors of \cite{Popescu2005, Kirchner2008, vazquez2011} developed techniques to expose resampling operations. Other contributions focused on the detection of the use of median filters \cite{Cao2010, Kirchner2010} or multiple image compressions \cite{Bianchi2011, Thai2016, mandelli2018multiple}. 
Furthermore, many researches aimed at spatially localizing traces left by editing operations applied locally on the image (i.e., splicing localization). 
A few examples of local image splicing are the insertion of a portion of an image into another one, or the deletion of a pixel area from the sample under attack.
To spot the tampering traces, 
many works in the literature
rely on the information carried by the so-called noise residual. This is a picture obtained by removing the high-level semantic content from the image \cite{Lyu2014, Cozzolino2015,Cozzolino2016}, for instance through high-pass filtering. 
In the last years, 
thanks to the automatic extraction of forensic traces executed by data-driven methods, techniques coming from the deep learning area have gained a lot of popularity in the forensics field. Especially \glspl{cnn}
have been exhaustively explored for the task of image tampering localization.
For instance, 
\cite{Bondi2017a} proposes a framework for image splicing localization by exploiting \gls{cnn}-based descriptors developed for camera model identification. 
Interestingly, some works combine \glspl{cnn} with the idea of noise residuals: they suppress the high-level image content by using either fixed \cite{Rao2016, Liu2018} or learned \cite{Bayar2016, Bayar2017} high-pass convolutional filters.  
More recent contributions further elaborate on this idea, providing tools that greatly improved the state-of-the-art performances on the splicing localization task \cite{Cozzolino2020}.

In addition to classical digital photographs, overhead imagery is recently becoming more accessible than before.
This is probably due to the increased availability of satellites equipped with imaging sensors and the widespread diffusion of public websites \cite{freesat} sharing this kind of images. This imagery represents data in a wide variety of modality, from optical (e.g., panchromatic, RGB), to thermal and \gls{sar} as well.

Despite the great availability of tools and techniques for the integrity analysis of natural images, the potentially malicious editing of overhead images is a growing concern. 
Indeed, as any other type of digital imagery, overhead data can be easily manipulated 
through editing software suites (e.g., Photoshop, GIMP, etc.) as well as through synthetic image generation tools
\cite{Abady2020, Zhao2021}, and examples of malicious modifications have been worrying the public opinion and media \cite{bbc, russia}. 
Unfortunately, the footprints characterizing the overhead image life-cycle are different from those of digital photographs.
Therefore, developing techniques to localize potential manipulations applied to overhead imagery is becoming a task of paramount importance. 

These reasons motivated the forensics community to develop techniques specifically tailored to overhead imagery, as state-of-the-art methods suited for digital photographs are likely bound to perform poorly if blindly applied to them. 
In this vein, the authors of \cite{Ho2005} propose
a
method based on a watermarking technique to detect doctored image regions in overhead imagery. 
In \cite{Luqman2017}, the authors propose a deep learning method for detecting inpainting attacks, while \cite{Yarlagadda2018} shows a tool for the localization of general overhead image manipulation combining a \gls{gan} with a one-class \gls{svm}. 
The authors of \cite{bartusiak2019splicing} rely on \glspl{gan} as well for detecting and localizing RGB image forgeries. 
In \cite{horvath2019anomaly}, the forgery detection problem on RGB data is formulated as an anomaly detection one. 
Similarly, in \cite{danipixelcnn}, splicing attacks are localized as deviations of the image pixel values from pristine distributions, in this case using generative auto-regressive models.

While the effort put by the multimedia forensics community is remarkable, to the best of our knowledge the problem of forgery localization on \gls{sar} imagery has never been investigated in the literature. However, since \gls{sar} products, especially those based on amplitude only, are easy to handle and modify even without specific expertise, their possible manipulation by malicious users is concerning.

This paper investigates the problem of splicing localization in amplitude \gls{sar} images. 
Specifically, we consider the situation in which a region of an amplitude \gls{sar} image has been substituted with another region coming from a different image, and some editing might have been applied to hinder this manipulation.

Our goal is to localize the manipulated region (i.e., performing splicing localization), providing a binary mask highlighting the manipulated area.
To do so, we rely on \glspl{cnn} to first extract a fingerprint reporting information on the forensic traces found in the analyzed image. 
The fingerprint extraction stage is inspired by existing state-of-the-art multimedia forensic methods, but we reformulate it to best suit the context of \gls{sar} imagery. 
Then, we exploit the extracted fingerprint to generate a tampering mask showing which pixels have undergone splicing. We propose three different methods, one supervised approach relying on \glspl{cnn} and two unsupervised approaches leveraging clustering techniques. 

To validate our findings, we construct a custom dataset of spliced amplitude \gls{sar} images by applying forgeries of different size and considering various editing operations performed on the manipulated data.
We compare with state-of-the-art algorithms for splicing localization on natural images, always achieving better localization performances.
Our results suggest that the forensic analysis on manipulated amplitude \gls{sar} images is feasible, as long as the splicing localization is performed being aware of the distinct nature of \gls{sar} imagery with respect to natural photographs.

The rest of the paper is organized as follows.
In Section~\ref{sec:background}, we provide some useful background on the deep learning tools employed in our proposed splicing localization pipeline, and on the \gls{sar} imagery generation process.
In Section~\ref{sec:problem}, we formulate the splicing localization problem on amplitude \gls{sar} images.
In Section~\ref{sec:method}, we illustrate our proposed method in details.
In Section~\ref{sec:experiments}, we provide all the information regarding the setup used for our experiments.
In Section~\ref{sec:results}, we discuss our experimental findings.
Finally, in Section~\ref{sec:conclusions}, we draw the final considerations on our work.

%% file: sec_background.tex
\section{Background}
\label{sec:background}
In this section, we provide the reader with some useful background information on the deep learning tools employed in this paper and on \gls{sar} imaging.

\subsection{Deep Learning tools}
\label{subsec:dl}
Deep learning is a prosperous study field that greatly improved state-of-the-art solutions for a wide range of applications, including multimedia forensics \cite{Verdoliva2018} as well as \gls{sar} image classification \cite{Geng2015} and automatic target recognition \cite{Chen2016}.
Among the most used deep learning tools, \glspl{cnn} had a great success as they proved handy in managing data with an intrinsic regular grid structure.
This is the case of digital images as well as remote sensing data, which are stored as multi-dimensional arrays.

In a nutshell, a \gls{cnn} can be seen as an operator that applies a series of parametric functions (e.g., linear filtering, non-linear saturation, matrix multiplications, etc.) to its input, in order to obtain a processed output.
The parameters of the applied functions are optimized (or ``learned'') during a preliminary stage called ``training''.
Depending on the task, the output of the \gls{cnn} can be a label (e.g., for classification problems), a heatmap (e.g., for segmentation or localization tasks), or any other processed version of the input (e.g., for denoising purpose).

In the following sections we illustrate some applications of these tools related to the use we make of them in our proposed method.

\subsubsection{Denoising Forensics}
\glspl{cnn} are being more and more exploited in the forensics field in the last few years 
\cite{Amerini2021}.
In this work, we are particularly interested in the use of the \gls{dncnn} \cite{Zhang2017}, a \gls{cnn} developed for image denoising that has been successfully exploited for forensics tasks \cite{Bonettini2018, Cozzolino2020}.
For instance, the authors of \cite{Cozzolino2020} proposed to use a \gls{dncnn} to extract the so-called Noiseprint.
This is a noise-like pattern that suppresses the vast majority of the image content and exposes editing-related artifacts due to local image forgery.
To extract the Noiseprint, the authors employ a particular training procedure which can be roughly summarized as follows:
\begin{enumerate}
    \item Consider a dataset of images coming from different devices.
    \item Apply the \gls{dncnn} to patches extracted from different images to obtain a series of noise patterns.
    \item Keep training the \gls{dncnn} to extract similar patterns for patches coming from the same pixel region (e.g., top-left, bottom-right, etc.) of the same device, and different patterns from patches coming from different regions and/or devices. 
\end{enumerate}
The last constraint is motivated by the idea of exploiting the spatial periodicity of camera-related artifacts, so that operations like image shift or rotation can be easily detected \cite{Cozzolino2020}.
In the end, the trained \gls{dncnn} is able to extract a noise-like heatmap that,
when analyzing pristine images, is self-consistent, whereas in case of spliced images clearly highlights the edited regions. 
This solution achieves state-of-the-art results on many image forensics datasets, and also proves useful in the analysis of remote sensing images, in particular overhead RGB data \cite{Cozzolino2020}.


\subsubsection{Segmentation Forensics}
A wide variety of \glspl{cnn}-based methods have been proposed for the task of image segmentation \cite{Minaee2021}.
A notable example due to its simplicity and accuracy is that of the \unet{} \cite{RFB15a}.
This network is characterized by a ``U'' shaped architecture.
This is made by a contracting path, i.e., a series of convolutional layers each one followed by pooling operations, and an expanding path specular to the contracting one but containing upsampling operator rather than pooling.
Skip connections are employed to concatenate and combine the output of the contracting path layers with the input of the mirrored layers of the expanding path.

In addition to image segmentation, the \unet{} has found a general appreciation also in the forensics field.
The authors of \cite{Kniaz2019} used a \unet-like architecture to train a \gls{gan} to hinder tampering artifacts in spliced images.
The authors of \cite{Bi2019} combined the idea of residual connections and \unet{} architecture to realize an end-to-end trainable network able not only to detect image manipulation attacks, but also to localize them precisely.

\subsection{SAR Imaging}
\label{subsec:sar_imaging}

\gls{sar} imagery has been widely adopted for a variety of tasks thanks to its characteristics of providing 
high-resolution images independently from cloud coverage, weather conditions and daylight \cite{Tomiyasu1978, Henderson1998, Oliver2004}.
Earth monitoring, 2D and 3D Earth surface mapping and change detection are just few examples of successful exploitation of \gls{sar} data \cite{Moreira2013}.

A \gls{sar} system is an imaging radar mounted on a platform moving in one direction (e.g., a satellite, an aircraft, etc.).
While moving, the system emits sequential high power electromagnetic waves through its antenna.
Waves interact with the objects they hit (i.e., the Earth surface) and are backscattered with modified amplitude and phase according to objects permittivity and physical properties (e.g., geometry, roughness). 
The antenna then collects these backscattered echoes that can be processed for the \gls{sar} image formation. 

\begin{figure}[t]
    \centering
    \includegraphics[width=.7\columnwidth]{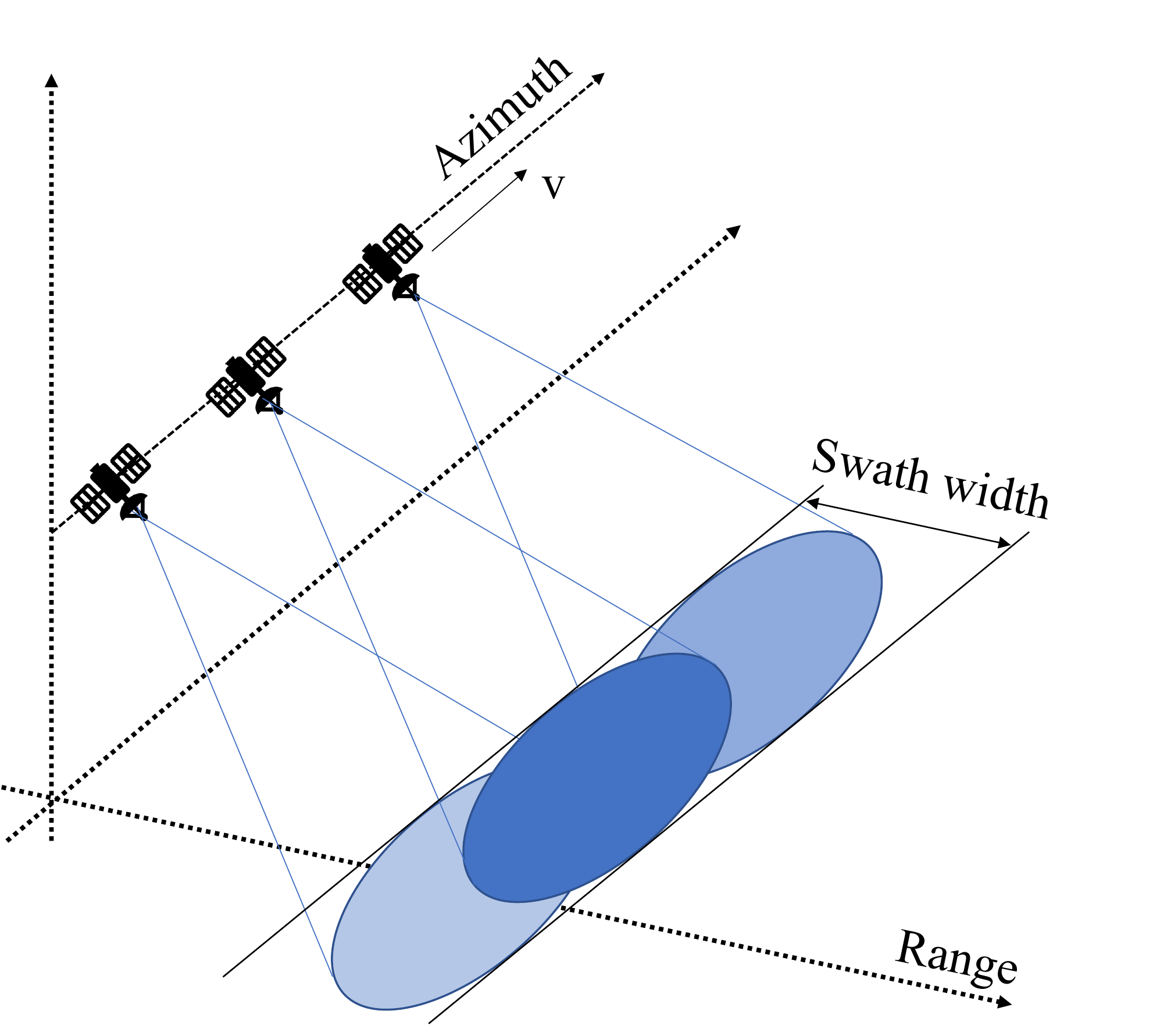}
    \caption{Simplified illustration of a \gls{sar} image acquisition. As the platform moves in the azimuth direction with velocity $\mathrm{v}$, the antenna emits electromagnetic pulses, receiving backscattered echoes. The pulse-echo collection process senses a portion of the Earth surface (a swath) in the azimut-range coordinates. 
    }
    \label{fig:sar_coordinates}
\end{figure}

A simplified schematic representation of this process is provided in Fig.~\ref{fig:sar_coordinates}.
The coordinates of \gls{sar} data are related to the motion of the platform at acquisition time.
As we can see in Fig.~\ref{fig:sar_coordinates}, the first dimension corresponds to the range (or fast time), i.e., the direction perpendicular to platform flight along which the electromagnetic beam travels.
The second dimension corresponds instead to the azimuth (or slow time), which is the actual trajectory of the platform.

\gls{sar} systems use frequency modulated pulse waveforms called chirps.
Chirps are characterized by constant amplitude and instantaneous frequency that is linearly modulated over time.
Depending on the application, different frequency bands are used for modulation, with the most popular being L (i.e., from \SI{1}{\giga\hertz} to \SI{2}{\giga\hertz}), C (i.e., from \SI{3.75}{\giga\hertz} to \SI{7.5}{\giga\hertz}) and X (i.e., from \SI{7.5}{\giga\hertz} to \SI{12}{\giga\hertz}) \cite{Moreira2013}.

Differently from optical sensors, data coming from echo signals is not interpretable as it is.
Additional processing called focusing (i.e., a double convolution both in the range and azimuth directions) is needed to obtain a visually interpretable image \cite{Moreira2013}.
The resulting \gls{sar} image is a complex 2D matrix, usually displayed in terms of intensity so that pixel values approximate the reflectivity of points on the ground.
This 2D matrix can be further processed, and different kinds of processing determine the existence of different so-called \gls{sar} products \cite{Oliver2004}.
As an example, this can be done to ensure calibration (i.e., each pixel value represents the correct value of reflectivity) and geocoding (i.e., associate the location of each pixel with a position on the ground).

Nowadays, a wide range of \gls{sar} products can be downloaded from online platforms.
Among these platforms, the Copernicus Open Access Hub \cite{copernicushub} is the online portal provided by the European Space Agency for downloading Copernicus Sentinel-1 Mission products.
The Sentinel-1 Mission products are generated according to different acquisition modes. The simplest one, the Stripmap mode, senses single continuous strips of Earth surface with a fixed antenna pattern (as Fig.~\ref{fig:sar_coordinates} depicts). Other acquisition modes instead acquire more than one measurement: for instance, the \gls{iw} emits three different pulses steering the antenna in the azimuth direction \cite{topsar}. This operation results in the generation of three different complex images (i.e., one per pulse in the azimuth direction), or sub-swaths, provided altogether in the \gls{sar} product.
\begin{figure}[t]
\centering
\includegraphics[width=.85\columnwidth]{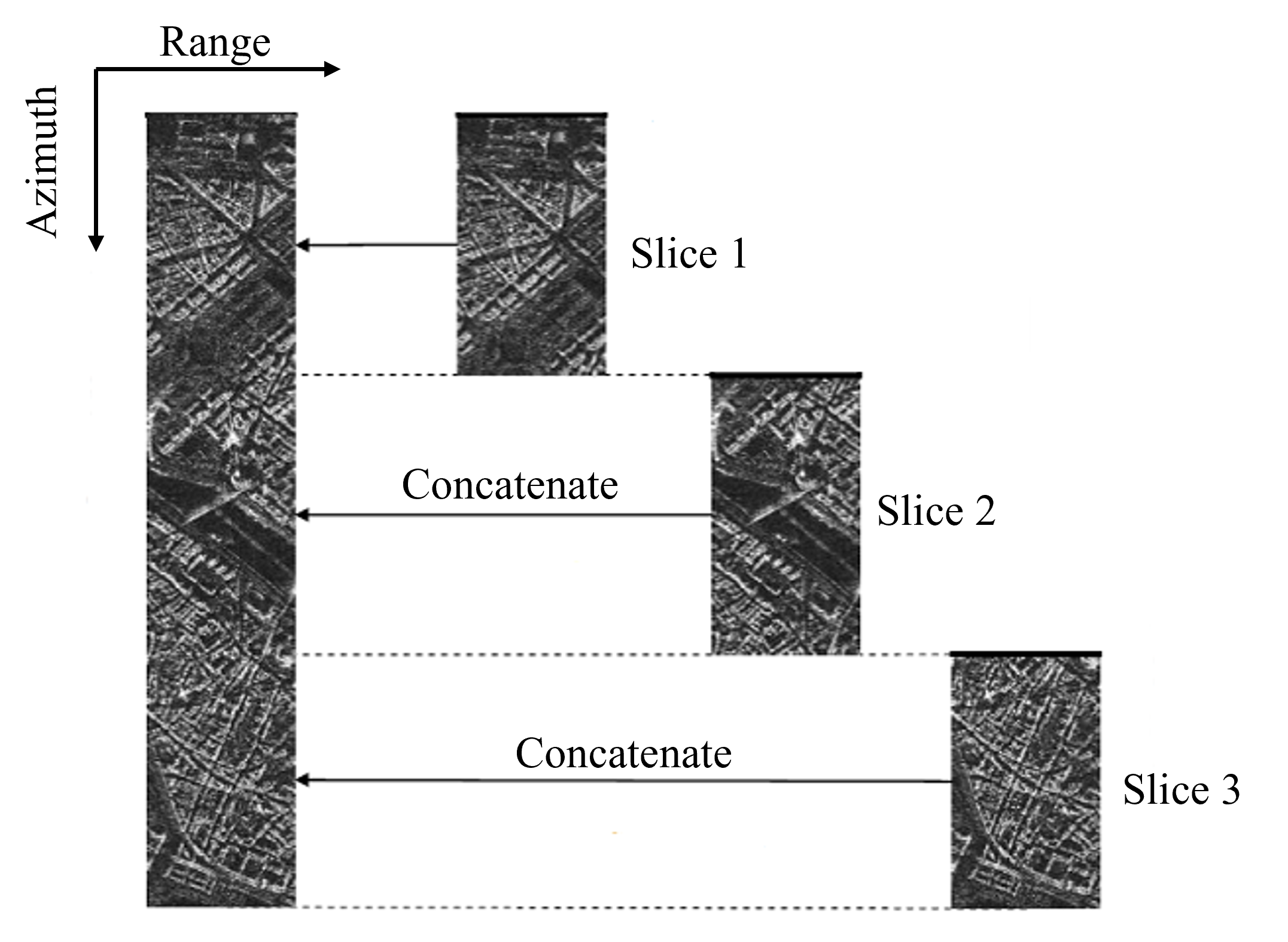}
\caption{Example of product slicing for a \gls{grd} product. After each slice has been resampled to a common grid, measurements are concatenated along the azimuth direction.}
\label{fig:product_slicing}
\end{figure}
\begin{figure}[t]
\centering
\includegraphics[width=\columnwidth]{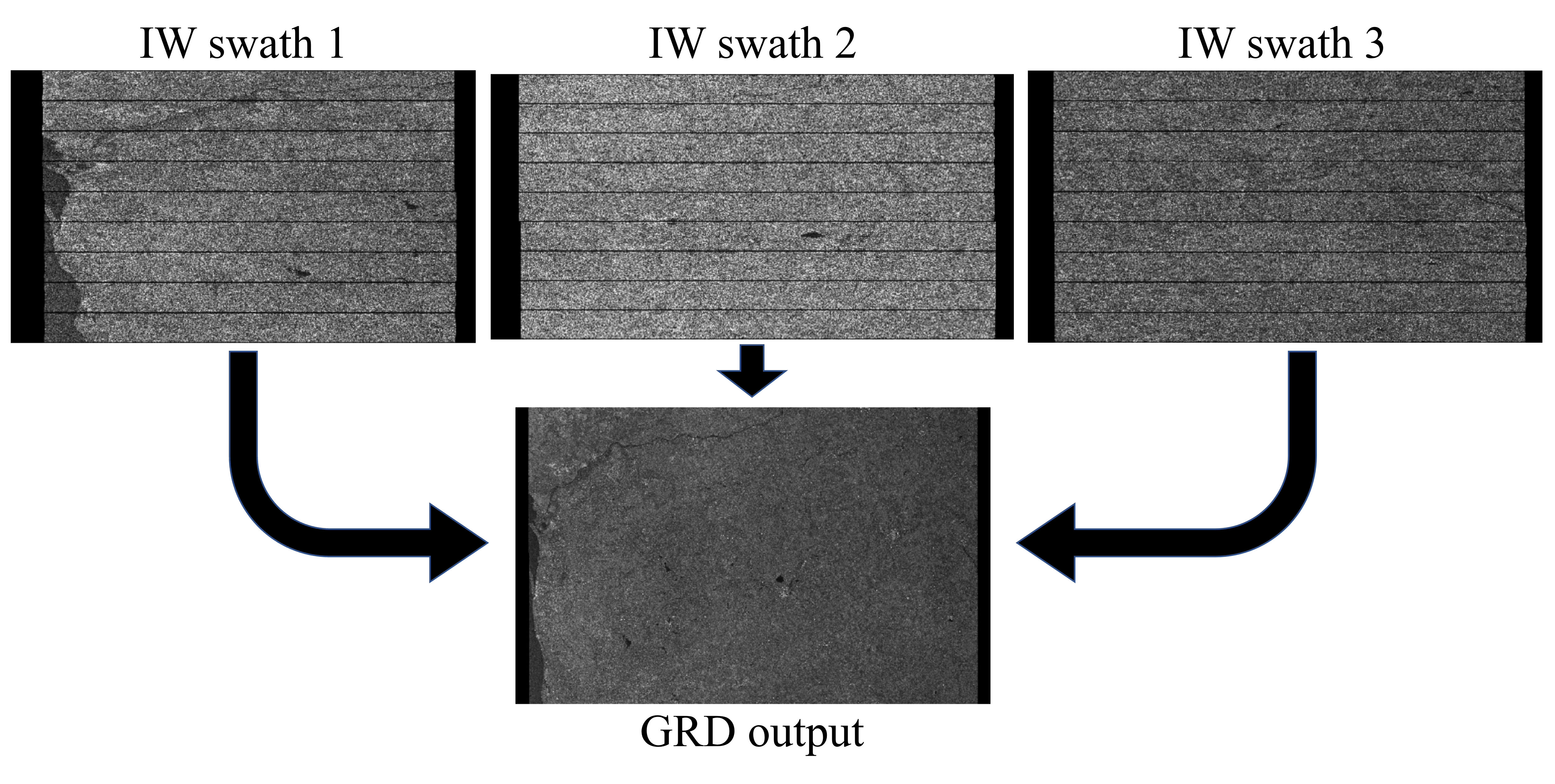}
\caption{Example of sub-swaths acquired with \gls{iw}, and of a full \gls{iw} acquired \gls{grd} product. For each sub-swath, only the magnitude of the complex signal has been plotted. 
The generation of a \gls{grd} product needs a resizing of all sub-swaths in order to have them coherently merged into a single image. 
}
\label{fig:GRD_merging}
\end{figure}

\gls{sar} products usually depict very large geographic areas. Many companies allow users to select an area of interest to be imaged by their systems. To do so, as the Earth surface coverage of a single echo is often insufficient, multiple signals are collected and
concatenated in a single continuous image. This is also the case of the Sentinel-1 Mission, where this operation is denoted as product slicing \cite{productslicing}. Fig.~\ref{fig:product_slicing} provides a graphical representation of it. Product slicing needs a resizing of all slices to a common grid in order to concatenate them.

Among the different Sentinel-1 products and acquisition modes, \gls{grd} products are probably the most common and accessible for a direct inspection. 
Indeed, they present scene reflectivity in ground-range coordinates, which are the azimuth-range coordinates projected on the Earth ellipsoid model \cite{wgs84}. 
This transformation allows to reduce the range geometric distortion and have each pixel placed in the correct position with respect to a reference plane \cite{groundrange}.
Moreover, in case multiple sub-swaths are available, all the available signals are fused together to obtain a single continuous image. This is the case of \gls{iw} acquired products. Fig.~\ref{fig:GRD_merging} provides an example of such process, which is based on a resizing pipeline \cite{sentel1userguide}. Finally, \gls{grd} products represent detected amplitude only, without bringing any phase information with them.
All these elements make \gls{grd} products easy to handle, but also easy to be manipulated with common image editing software tools.

%% file: sec_problem.tex
\section{Problem formulation}\label{sec:problem}

\gls{sar} products differ from natural photographs for a variety of reasons.
For instance, the concept of single shot is hardly defined.
Indeed, \gls{sar} signals are continuously acquired through moving sensors.
Individual products are then generated for manageability reasons by merging several acquisitions.
Moreover, some \gls{sar} products like \gls{grd} are obtained through a very specific chain of operations that has nothing in common with the ones usually employed in natural photography.

However, from the perspective of an end-user with no specific experience on overhead imagery, amplitude \gls{sar} products can be considered close to natural photographs when it comes to their manipulation.
Indeed, as long as they are provided in single polarization, since they present amplitude information only they can be processed as a matrix of real numbers with any common image editing tool. 
This is the case of \gls{grd} products for instance.

Since \gls{grd} products, especially those acquired in \gls{iw} mode, are popular \cite{Schmitt2018, Schmitt2019} yet easy to be manipulated, it is reasonable to consider them a vulnerable asset from a forensics perspective.
Given these premises, in this work we focus on \gls{grd} products.
Specifically, we consider images derived from \gls{grd} products in C-band in single vertical polarization, all acquired in \gls{iw} mode. From now on, we refer to them as \gls{grd} images, or \gls{grd} tiles as they are typically mentioned in the overhead field.


In this work, we are interested in assessing the integrity of a \gls{grd} image tile at a local level and at a small granularity.
In particular, given a manipulated tile,
we want to 
localize
which pixels have been affected by the editing.
As manipulation we consider image splicing attacks, i.e., the insertion in a target tile of a portion coming from a different source tile.
Moreover, we consider that the target region may have undergone optional editing with image processing operations (e.g., blurring, resizing, noise addition, etc.) in order to render the attack more credible and visually appealing.
For instance, a resizing might be needed to match the source and target tile resolution and avoid making the splicing easily detectable at visual inspection.

More formally, we define the coordinates of a pixel of a $U \times V$ resolution tile as $(u, v)$, where $u \in [1, \dots, U]$ and $ v \in [1, \dots, V]$.
$U, V$ are the number of pixels per row and column, respectively.
Let $\mathbf{T}_D$ and $\mathbf{T}_T$ be two pristine tiles.
$\mathbf{T}_D$ is the donor tile, whereas $\mathbf{T}_T$ is the target tile.
Defining 
$\mathcal{S}$ as the region of $\mathbf{T}_T$ under splicing attack, 
the resulting spliced tile $\mathbf{T}_S$ is defined as:
\begin{equation}
    \label{eq:splicing_def}
    \mathbf{T}_S(u, v) = 
    \begin{cases}
    e\bigl(\mathbf{T}_D)(u, v), &\quad \text{if $(u,v)\in\mathcal{S}$}\\
    \mathbf{T}_T(u, v), &\quad \text{if $(u,v)\notin\mathcal{S}$}
    \end{cases},
\end{equation}
with $e(\cdot)$ being a suitable editing function (e.g., blurring, resizing, noise addition, rotation, shearing, affine transforms, etc.).

\begin{figure}[t]
    \centering
    \includegraphics[width=1\columnwidth]{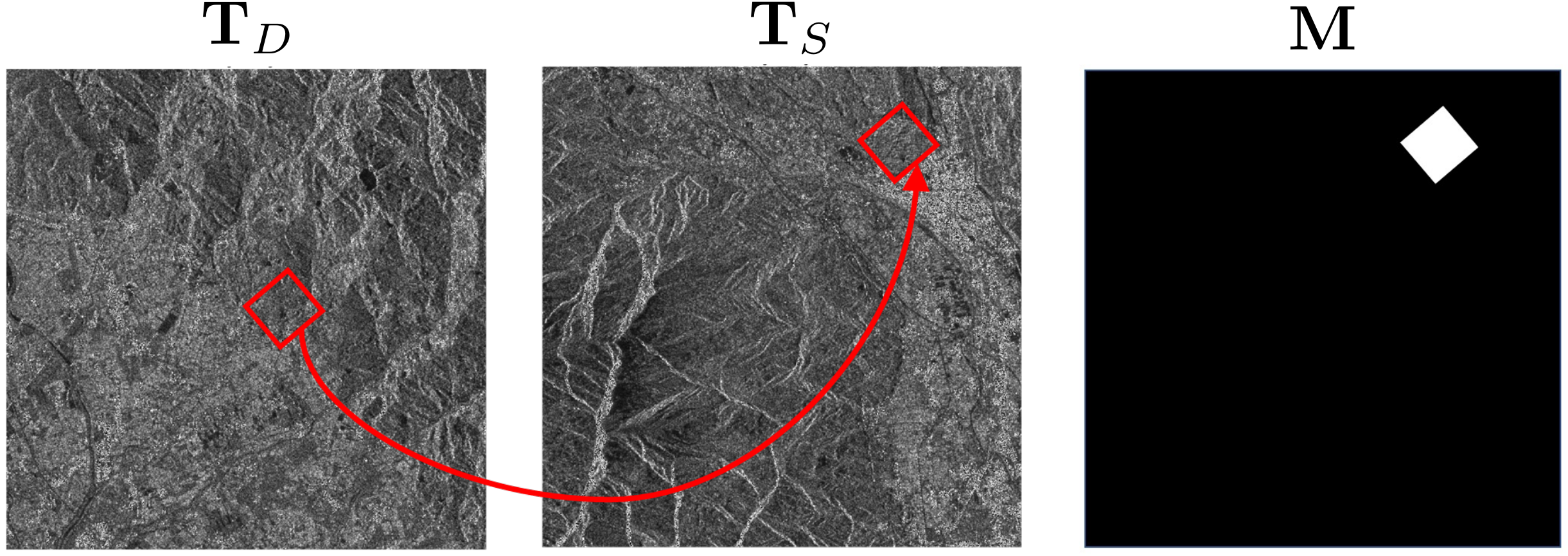}
    \caption{Example of a splicing operation, with donor tile $\mathbf{T}_D$ on the left, spliced tile $\mathbf{T}_S$ at the center and tampering mask $\mathbf{M}$ at the right. 
    In the spliced tile, the outskirts of a urban area are covered.}
    \label{fig:splicing_example}
\end{figure}

The pixel-by-pixel integrity of the tile $\mathbf{T}_S$ can be represented by a tampering mask $\mathbf{M}$ with the same resolution of $\mathbf{T}_S$, where each pixel takes a binary value 0 or 1 depending on the pixel being pristine or manipulated, respectively.
Formally, the tampering mask $\mathbf{M}$ has pixel values equal to
\begin{equation}
    \mathbf{M}(u, v) = \begin{cases}
    1,\quad \text{if $(u,v)\in\mathcal{S}$}\\
    0,\quad \mathrm{otherwise}
    \end{cases}.
    \label{eq:mask_def}
\end{equation}
The goal of this paper is the localization of the spliced region $\mathcal{S}$ by estimating a tampering mask $\mathbf{\hat{M}}$ as close as possible to $\mathbf{M}$ from the sole analysis of the tile $\mathbf{T}_S$.
Fig.~\ref{fig:splicing_example} provides a graphical representation of the splicing operation together with a tampering mask $\mathbf{M}$.

%% file: sec_method.tex
\section{Amplitude SAR image splicing localization}
\label{sec:method}

In the forensics literature, it is well known that both the acquisition device and processing operations leave peculiar traces on digital photographs.
These traces can be exploited to expose forgeries \cite{Lukas2006, Cozzolino2020}.
As the considered amplitude \gls{sar} products undergo a wide variety of operations from their acquisition to the final production (e.g., re-sampling, de-ramping, ground-range projection, etc.), it is reasonable to assume that different products may contain different processing traces.
Due to the nature itself of the \gls{sar} signal and of the non-linear operations employed, even amplitude \gls{sar} products coming from the same satellite might present different traces relative to the processing executed for generating them.

Leveraging this idea, we propose a splicing localization method that exposes and highlights inconsistencies in the analyzed spliced tile $\mathbf{T}_S$ due to the different processing that target and donor tiles have undergone, and to any editing trace left by the attacker in the splicing operation.
This is done by extracting a fingerprint inspired by Noiseprint \cite{Cozzolino2020} from the tile under analysis.
This fingerprint, which suffices in spotting at a visual inspection the spliced region $\mathcal{S}$, is then further processed to estimate a binary tampering mask $\mathbf{\hat{M}}$ as close as possible to $\mathbf{M}$.

\begin{figure}[t]
\centering
\includegraphics[width=0.9\columnwidth]{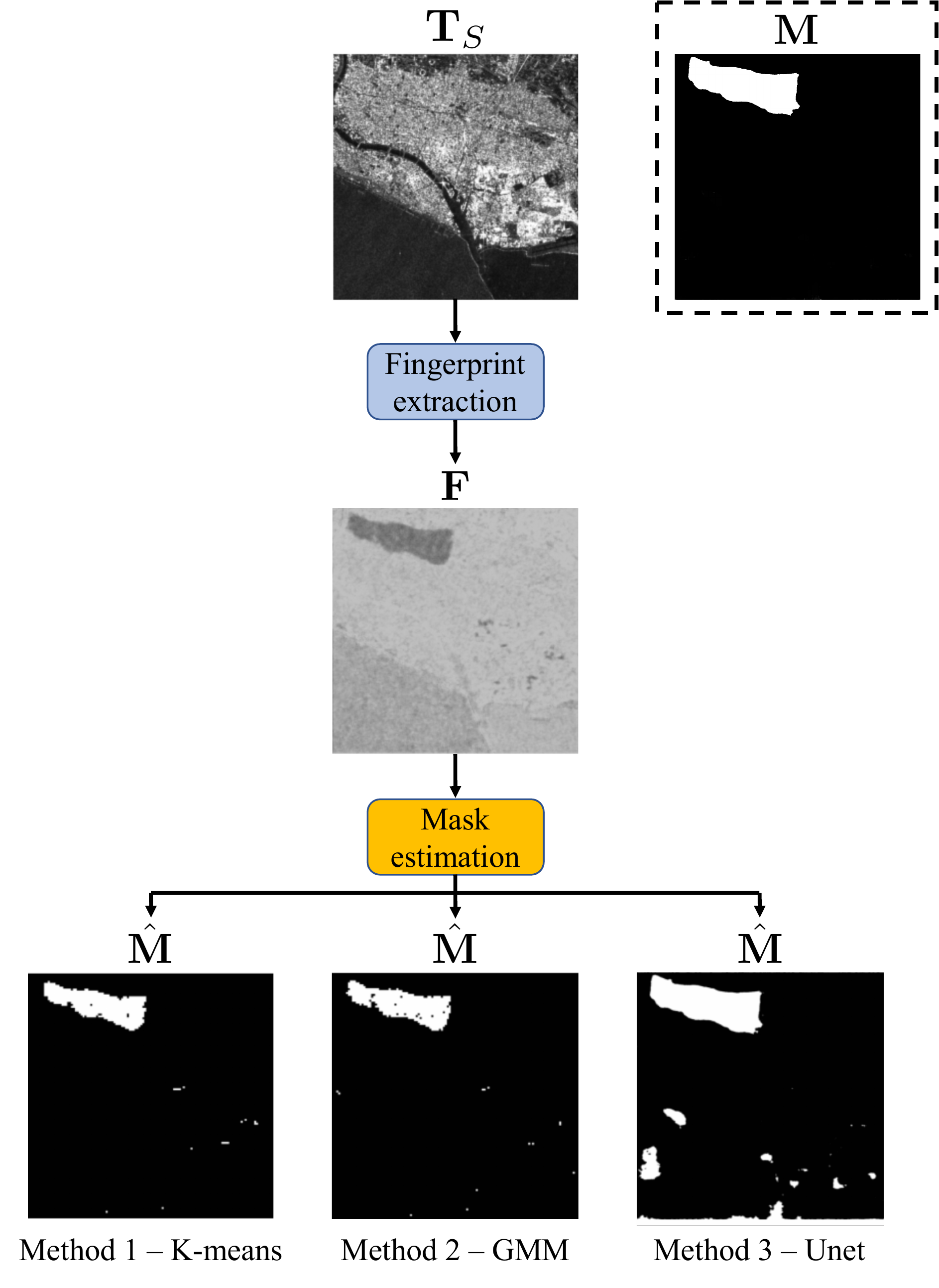}
\caption{Schematic illustration of the proposed processing pipeline. A fingerprint $\mathbf{F}$ is extracted from the spliced tile $\mathbf{T}_S$ under investigation. Then, a binary tampering mask $\mathbf{\hat{M}}$ can be estimated through three different methods. For the sake of clarity, we also report the ground-truth tampering mask $\mathbf{M}$.
}
\label{processing_pipeline}
\end{figure}

To summarize, our splicing localization process follows a two-stage pipeline (see Fig.~\ref{processing_pipeline}):
\begin{enumerate}
    \item \textbf{Fingerprint extraction} - Using a properly designed \gls{cnn}, a fingerprint $\mathbf{F}$ with the same resolution of $\mathbf{T}_S$ highlighting any local inconsistencies due to splicing attacks is obtained.
    \item \textbf{Tampering mask estimation} - Starting from the fingerprint $\mathbf{F}$, using either unsupervised or supervised approaches, 
    a tampering mask $\mathbf{\hat{M}}$ is estimated.
\end{enumerate}
In the following, we provide additional details about each step of the proposed method.

\subsection{Fingerprint Extraction}
\label{subsec:fingerprint}
The goal of this step is the extraction of a fingerprint $\mathbf{F}$ that visually highlights the spliced region $\mathcal{S}$ in an analyzed tile.
To do so, we leverage the recent forensics literature.
In particular, the Noiseprint \cite{Cozzolino2020} method 
shows promising results in highlighting editing traces even on data distant from natural photographs.

For our fingerprint extractor, we exploit the characterization capability offered by Noiseprint and further adapt it to the context of \gls{sar} imagery.
In particular, given a spliced tile $\mathbf{T}_S$, we 
extract a fingerprint $\mathbf{F}$ with the same pixel resolution.
Our goal is to make this fingerprint clearly highlighting spliced regions just by visual inspection. Formally, we define $\mathbf{F}$ as:
\begin{equation}
    \label{eq:fingerprint}
    \mathbf{F} = f\bigl(\mathbf{T}_S\bigr), 
\end{equation}
where $f(\cdot)$ represents the fingerprint extraction operator, i.e., the \gls{dncnn} network after it has been trained.

\begin{figure}[t]
\centering
\includegraphics[width=\columnwidth]{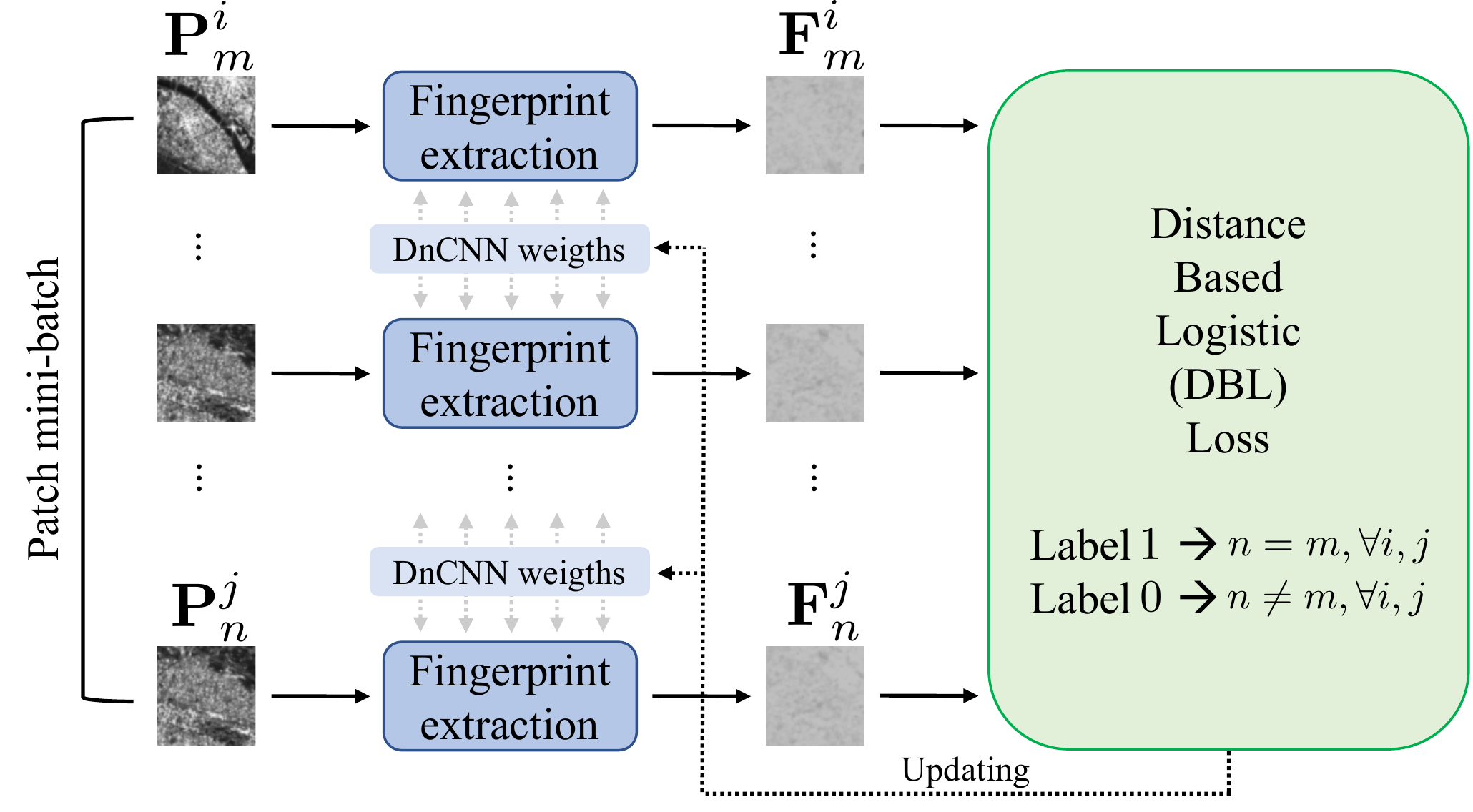}
\caption{Training procedure of the fingerprint extraction block. For each patch in the mini-batch, we extract a fingerprint by means of the \gls{dncnn}. We iteratively update the \gls{dncnn} weights by comparing all the fingerprints of the mini-batch in a pair-wise fashion through the \gls{dbl} loss. Pairs of fingerprints extracted from the same amplitude \gls{sar} product are associated with label $1$, otherwise label $0$ is assigned. }
\label{fig:training_fingeprint_extraction}
\end{figure}

For the \gls{dncnn} training, we adopt the following pipeline:
\begin{enumerate}
    \item We collect a number of tiles coming from $M$ different amplitude \gls{sar} products, all generated by the same satellite. These tiles are pristine, i.e., they have not been tampered with in any way.
    \item From the tiles of each product, we extract a number of patches. 
    The $i$-th patch extracted from the tiles of the $m$-th amplitude \gls{sar} product is referred to as $\mathbf{P}^i_{m}$.
    \item From each patch $\mathbf{P}^i_{m}$, we extract the related fingerprint $\mathbf{F}^i_{m} = f(\mathbf{P}^i_{m})$.
    \item We iteratively update the \gls{dncnn} weights by processing small batches of patches. In particular, 
    given a mini-batch of patches,
    we process their extracted fingerprints 
    with the \gls{dbl} loss presented in \cite{distancebasedlogisticloss}. 
    This loss function computes the pairwise squared Euclidean distance between all the analyzed fingerprints. 
    The objective is to make the fingerprints self-consistent 
    if and only if they are extracted from the same amplitude product.
    Consistent fingerprint pairs (i.e., coming from the same \gls{sar} product) are associated with a desired low value for the Euclidean distance, while non consistent fingerprint pairs (i.e., coming from different products) are associated with a desired high Euclidean distance.
    More formally, we define the fingerprint pair $(\mathbf{F}^i_{m}, \mathbf{F}^j_{n})$ as consistent if $n = m, \forall i, j$. The fingerprint pair is non consistent if $n \neq  m, \forall i, j$.
    To do so, we assign a label $1$ to all consistent pairs of fingerprints and label $0$ otherwise.
    \item We process the training patches by continuously updating the \gls{dncnn} weights until we reach some desired performance metrics.
\end{enumerate}
For clarity's sake, Fig.~\ref{fig:training_fingeprint_extraction} depicts a sketch of the training pipeline of the proposed fingerprint extractor.

When training is finished, $f(\cdot)$ implements the desired fingerprint extraction function defined in \eqref{eq:fingerprint}. This function allows to extract a fingerprint $\mathbf{F}$ that captures traces relative to the processing pipeline of the acquired product, highlighting splicing attacks as inconsistencies in these traces. 
It is worth noticing that $f(\cdot)$ scales with the input resolution, so that tiles of different pixel dimensions can be processed seamlessly.

With respect to the original training procedure outlined in Section~\ref{sec:background}, our pipeline has been modified taking into consideration the differences between natural images and \gls{sar} products. For instance, the presence of coherent artifacts in specific positions of the grid of pixels can be hardly demonstrated for amplitude \gls{sar} products. 
On top of that, pixels in natural images are temporally coherent: since they are approximately acquired at the same time instant, the signal they represent has basically no spatial nor temporal discontinuities.
This is unfortunately not true for \gls{sar} products. As a matter of fact, we have seen  that \gls{sar} images are generated concatenating different measurements. This implies that operations such as product splicing, together with other processing characteristic of amplitude \gls{sar} products, 
might alter and hinder the presence of generation artifacts with a regular spatial distribution. 

We summarize our main elements of difference with respect to \cite{Cozzolino2020} as follows:
\begin{enumerate}
    \item Noiseprint requires the collection of images coming from different devices (i.e., individual cameras). We exploit instead a number of tiles from $M$ different amplitude \gls{sar} products all coming from the same satellite.
    Our assumption is that tiles
    of the same amplitude product underwent the same processing pipeline, whereas tiles of different products present different processing traces. 
    
    \item Noiseprint trains the \gls{dncnn} by comparing pairs of patches, 
    giving a positive label in the \gls{dbl} only if patches come from the same pixel region and device. 
    Reasonably this constraint should not hold for \gls{sar} images, thus we relax it and give a positive label whenever the patches come from the same amplitude product, regardless of the pixel region of extraction.
    \item Noiseprint does not employ any data-augmentation strategy during training.
    We propose to include resizing as data-augmentation.
    The reason behind this choice is twofold: on one hand, resizing might improve the extractor robustness to editing operations applied to hinder the localization of $\mathcal{S}$. On the other hand, 
    we have seen in Section~\ref{sec:background} that each \gls{sar} product is characterized by a number of resizing operations leaving peculiar traces. For this reason, we apply resizing and propose to consider all resized tiles as coming from new \gls{sar} products.  
    This is a good solution to
    enlarge the number and variety of products at disposal. 
\end{enumerate}
Such modifications, even if at a first glance might appear negligible, will later show to improve the performances of the pipeline with respect to adopting the baseline training procedure as it is.

\begin{figure}[t]
\centering
\subfloat[Tile $\mathbf{T}_{S_1}$.]{\includegraphics[width=0.3\columnwidth]{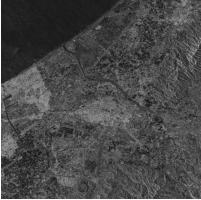}
\label{fig:fingerprint_image_1}}
\hfil
\subfloat[Estimated $\mathbf{F}_1$.]{\includegraphics[width=0.3\columnwidth]{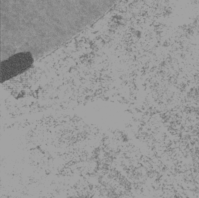}
\label{fig:fingerprint_fp_1}}
\hfil
\subfloat[Mask $\mathbf{M}_1$.]{\includegraphics[width=0.3\columnwidth]{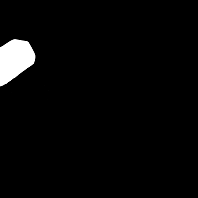}
\label{fig:fingerprint_mask_1}}

\subfloat[Tile $\mathbf{T}_{S_2}$.]{\includegraphics[width=0.3\columnwidth]{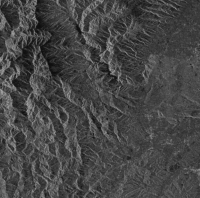}
\label{fig:fingerprint_image_2}}
\hfil
\subfloat[Estimated $\mathbf{F}_2$.]{\includegraphics[width=0.3\columnwidth]{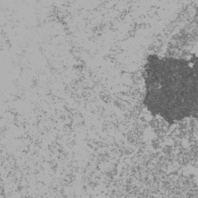}
\label{fig:fingerprint_fp_2}}
\hfil
\subfloat[Mask $\mathbf{M}_2$.]{\includegraphics[width=0.3\columnwidth]{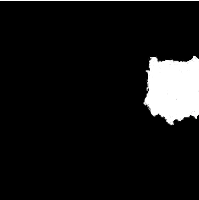}
\label{fig:fingerprint_mask_2}}

\caption{Examples of fingerprints extracted from spliced tiles. We can notice a spliced area highlighted in the fingerprints $\mathbf{F}_1$ and $\mathbf{F}_2$. The precise location of the spliced areas is reported in the tampering masks $\mathbf{M}_1$ and $\mathbf{M}_2$.}
\label{fig:fingerprint_examples}
\end{figure}

Fig.~\ref{fig:fingerprint_examples} reports some examples of spliced tiles, their tampering masks and the fingerprints extracted with function $f(\cdot)$.
As we can see, even though the fingerprints $\mathbf{F}_1$ and $\mathbf{F}_2$ are not binary images yet, spliced areas are easily recognizable at this stage of the pipeline already.

\subsection{Tampering Mask Estimation}
\label{subsec:mask_estimation}
Once the fingerprint $\mathbf{F}$ is extracted, the next step in the pipeline is tampering mask estimation.
This stage segments the fingerprint to generate a binary mask $\mathbf{\hat{M}}$ representing the integrity of the spliced tile $\mathbf{T}_S$ given as input.

Many forensics methods in the literature provide a binary heatmap decision highlighting the spliced region. The most common approaches rely on
automatic thresholding
or 
two-class clustering. 
However, such procedures may lead to a non-efficient binary partition of the feature space \cite{Hosseini2019}, as the presence of scene content 
might still appear in the generated tampering mask. 
Such phenomenon is noticeable also in the fingerprints shown
in Fig.~\ref{fig:fingerprint_examples}.
For instance, the fingerprint $\mathbf{F}_1$ presents some texture related to the sea and the urban areas surrounding $\mathcal{S}$.

Having a binary heatmap however is of paramount importance for supporting the work of forensic analysts. For this reason the second step of our pipeline is dedicated in providing the most detailed possible tampering mask $\mathbf{\hat{M}}$. To do so, we propose different methods that deeply analyze the fingerprint $\mathbf{F}$ to provide an efficient binary partition of it. 
To this end, we have considered three different techniques, which can be divided into two families:
\begin{itemize}
    \item \textbf{Unsupervised} approaches. With these techniques, 
    we first partition $\mathbf{F}$ into different clusters. Then, starting from these partitions, different candidate masks are compared to choose the most appropriate one. 
    
    We propose two unsupervised methods: the first one is based on the K-means clustering algorithm \cite{macqueen1967}; the second one is based on \glspl{gmm} \cite{ Mclachlan2000}. As unsupervised methods, both techniques do not require a preliminary stage of training.
    
    \item \textbf{Supervised} approaches. In this scenario, we estimate the tampering mask $\mathbf{\hat{M}}$ using classic \glspl{cnn} adopted in the image segmentation field. Specifically, we propose a supervised strategy based on the well-known \unet{} architecture \cite{RFB15a}. This method requires a preliminary stage of training.
\end{itemize}

\subsubsection{Unsupervised K-means-based mask estimation}

The K-means algorithm \cite{macqueen1967} is well known in the signal processing community and it has been historically used to perform clustering operations. Given a number of observations, the algorithm partitions them into a finite set of groups, called clusters, assigning each observation to the group showing the nearest mean distance (e.g., the Euclidean distance) 
from it. 

The assumption behind its use in our pipeline is that the spliced region $\mathcal{S}$ is well localized in the fingerprint $\mathbf{F}$. A good estimate of the tampering mask will reasonably present a well localized cluster of pixels.
In a nutshell, we propose to look for different clusters of pixels in $\mathbf{F}$, compare their compactness and then choose the most compact one to estimate the final tampering mask $\mathbf{\hat{M}}$.

To do so, we first divide $\mathbf{F}$ into a set of non-overlapping patches $\mathbf{P}_n$, $n=1, \dots, N$, with $N$ being the total number of patches in $\mathbf{F}$. These patches are the observations used by the K-means algorithm to cluster $\mathbf{F}$ based on their Euclidean distances.
After the algorithm converges, the fingerprint $\mathbf{F}$ is divided into $C$ clusters. We define the set of coordinates of pixels belonging to the $c$-th cluster as $\mathcal{P}_c = [\mathcal{U}_c, \mathcal{V}_c]$. 
$\mathcal{U}_c$ and $\mathcal{V}_c$ are the sets of row and column coordinates, respectively.

It is worth noticing that, if the pixels belonging to a cluster are close to each other, this might be indicative of the presence of a well localized spliced area in the fingerprint $\mathbf{F}$. We therefore need a measure of proximity of the pixels. 
As a metric, we propose to compute the variance of the coordinate values of the pixels belonging to each cluster. 
The smaller the variance, the more compact the cluster.
Thus, the best localized cluster can be estimated as:
\begin{equation}
    \label{eq:comp_metric}
    \hat{c}=\arg \min_{c} \mu\Bigl(\sigma^2\bigl(\mathcal{U}_c\bigr), \sigma^2\bigl(\mathcal{V}_c\bigr) \Bigr), \quad c=1, \dots, C,
\end{equation}
where $\sigma^2$ is the variance and $\mu$ is the arithmetic mean.

Starting from the pixel coordinates of $\mathcal{P}_{\hat{c}} = [\mathcal{U}_{\hat{c}}, \mathcal{V}_{\hat{c}}]$, i.e., the coordinates of the best localized cluster, we finally create a binary segmentation mask of the fingerprint $\mathbf{F}$.
We do so by assigning a positive label to all the pixels belonging to that cluster, and a negative label to all those not belonging to it. 
Following the convention introduced in \eqref{eq:mask_def}, we assign $1$ as positive label and $0$ as negative one. 
The final tampering mask $\mathbf{\hat{M}}$ can be formally defined as:
\begin{equation}
    \label{eq:k-means_mask}
    \mathbf{\hat{M}}(u, v) = \begin{cases}
    1,\quad \text{if $(u,v)\in\mathcal{P}_{\hat{c}}$}\\
    0,\quad \mathrm{otherwise}
    \end{cases}.
\end{equation}




\subsubsection{Unsupervised GMM-based mask estimation}

Mixture distributions are a powerful statistical tool consisting in the description of data by linearly combining basic distributions, such as Bernoulli, Dirilichet, or Gaussian \cite{Mclachlan2000}. They have been studied for years also for the task of data clustering, and especially \glspl{gmm} have proven extremely handy and simple to use adopting the \gls{em} algorithm \cite{Dempster1977}.

The basic functioning of the \gls{em} algorithm is not too dissimilar from the K-means. Given a set of observations, the \gls{em} fits a mixture of $C$ different Gaussian distributions to the data.
The mixture is such that each $c$-th component groups together observations that have been likely generated by the same Gaussian distribution.
This performs a clustering of the data based on the probability of each cluster having generated a data point.


Inside our pipeline, the use of \glspl{gmm} is really close to K-means.
In this case as well, we propose to divide the fingerprint $\mathbf{F}$ into non-overlapping patches $\mathbf{P}_n$, $n=1, \dots, N$, with $N$ being the total number of patches in $\mathbf{F}$. These are the observations used by the \gls{em} algorithm. 
After the algorithm converges, the fingerprint $\mathbf{F}$ is divided again into $C$ clusters, where the elements of the clusters are paired based on how well their values can be described by the same Gaussian distribution.
We then look for the most compact cluster $\hat{c}$, following the same methodology applied for the K-means mask generation method. Using \eqref{eq:comp_metric}, we look for the cluster whose pixels' coordinate values show the smallest variance. The final estimated tampering mask is defined as in \eqref{eq:k-means_mask}. 



\subsubsection{Supervised CNN-based mask estimation}


We propose a supervised mask estimation strategy relying on the U-Net architecture \cite{RFB15a}.
Our choice fell over this network as it is easy and fast to train, while achieving really competitive performances in the segmentation of a variety of imagery data, from \gls{sar} \cite{Wei2020}, to overhead RGB for road extraction \cite{Zhang2018}, to seismic images salt segmentation \cite{Alfarhan2020}, and of course medical imagery \cite{Siddique2021}.

In the context of our proposed method, the use of the \unet{} translates in using the network to generate a tampering mask estimate $\mathbf{\hat{M}}$ starting from an input fingerprint $\mathbf{F}$. 
The proposed method first extracts a probability mask $\mathbf{\hat{M}}^u$ defined as
\begin{equation}
    \label{eq:unet_mask_prob}
    \mathbf{\hat{M}}^u = u\bigl(\mathbf{F}\bigr), 
\end{equation}
where $u(\cdot)$ is the fingerprint segmentation function implemented by the \unet{}.
$\mathbf{\hat{M}}^u$ has the same pixel resolution of $\mathbf{F}$, and each pixel presents values close to 0 when there is a low probability for that pixel of being spliced, and values close to 1 when there is an high probability of manipulation. 

In order for the function $u(\cdot)$ to correctly implement a coherent segmentation, the deployment of the \unet{} needs a stage of training, in which the network learns to retain only the information regarding the localization of the spliced region $\mathcal{S}$.
Such training can be done with a dataset of $K$ spliced tiles $\mathbf{T}_{S_k}$, $k=1, \dots, K$. For each of them, a corresponding ground-truth tampering mask $\mathbf{M}_k$ together with an extracted fingerprint $\mathbf{F}_k$ needs to be provided.

During the training phase, every 
fingerprint $\mathbf{F}_k$ is processed according to \eqref{eq:unet_mask_prob} to obtain the tampering mask estimate $\mathbf{\hat{M}}^{u}_{k}$.
The network performances are then evaluated comparing the ground-truth tampering masks $\mathbf{M}_k$ and the \unet{}-estimates $\mathbf{\hat{M}}^{u}_{k}$. 
We propose to do so by minimizing the sum of Dice loss \cite{Deng2018} and Focal loss \cite{Lin2018}. 
For a complete definition and a more comprehensive discussion on both losses we refer the reader to the original papers.
For the sake of our discussion, it suffices to say that both have proven to help reducing the overly thick boundaries in the segmented objects that usually present when training segmentation networks with a simple binary cross-entropy loss \cite{Deng2018}.

At deployment stage,
we propose to impose a threshold $\tau$ on the pixel values of the estimated mask $\mathbf{\hat{M}}^u$ derived from a query fingerprint $\mathbf{F}$. 
The final estimated mask is equal to:
\begin{equation}
    \mathbf{\hat{M}}(u, v) = \begin{cases}
    1,\quad \text{if } \mathbf{\hat{M}}^u(u,v)\ge \tau\\
    0,\quad \mathrm{otherwise}
    \end{cases}.
    \label{eq:unet_masK_bin}
\end{equation}
It is worth noticing that the training of the fingerprint extractor (presented in Section~\ref{subsec:fingerprint}) and the training of the \unet{} for the forgery mask estimation do not simultaneously happen: we first need to train the fingerprint extractor to generate the fingerprint $\mathbf{F}$, and then, in a second stage, the \unet{}. 
Furthermore, 
notice that the fingerprint extractor is trained on pristine tiles only; the \unet{} instead needs to be trained on spliced tiles.


%% file: sec_setup.tex
\section{Experimental Setup}\label{sec:experiments}

In this section, we describe the details regarding our experimental setup, including the dataset collection procedure, the training strategy together with the hyperparameters for the \gls{cnn}-fingerprint extractor and mask estimation methods, and finally the metrics used for our method evaluation.

\subsection{Dataset}
\label{dataset}
As introduced in Section ~\ref{sec:method}, in our work we considered \gls{sar} \gls{grd} products in single vertical polarisation.
More specifically, we downloaded from the Copernicus Open Access Hub 20 products acquired in IW mode coming from the Sentinel-1 mission. All products have been sensed by the same satellite (S1-B), present high spatial resolution, and overall dimensions in pixels roughly around $20000\times 20000$.

Given the size of these acquisitions, each of them has been divided into non-overlapping tiles $\mathbf{T}$ $1024\times 1024$ pixels wide. 
These operation allowed us to work at a local level with small granularity and making the input easily processable by our networks.
From each product, we extracted $300-400$ tiles, resulting in a total of approximately $8000$ tiles.
These data constituted the basis for all the steps of our experiments.
Indeed, starting from these samples we managed to create the following datasets:
\begin{itemize}
    \item \textbf{Fingerprint Extraction Dataset (FED)}. This is the dataset  of pristine tiles used for training the fingerprint extraction function $f(\cdot)$ defined in Section~\ref{subsec:fingerprint};
    \item \textbf{Spliced Dataset 1 (SD1)}.  This is a dataset of spliced tiles used for training the \unet{} segmentation function $u(\cdot)$ defined in Section~\ref{subsec:mask_estimation};
    \item \textbf{Spliced Dataset 2 (SD2)}.This dataset is again constituted by spliced samples, but that have never been seen during the training nor validation of the \unet{}. We used these tiles to test the performances of the complete pipeline with all its tampering mask estimation methods.
\end{itemize}
Table \ref{tab:dataset_composition} reports a summary of the different datasets used in the paper.
In the following paragraphs, we provide further details on the creation and usage of each set. 

\begin{table}[t]
\centering
\caption{Dataset composition and use of each set of tiles.}
\label{tab:dataset_composition}
\resizebox{\columnwidth}{!}{
\begin{tabular}{@{}cccccc@{}}
\toprule
\textbf{Set} & \multicolumn{2}{c}{\textbf{\# of tiles}} & \multicolumn{2}{c}{\textbf{Training}} & \textbf{Testing} \\ & & &
\textit{Fingerprint} & \textit{U-Net}\\ 
\midrule
FED & Pristine tiles & $4000$ & \checkmark & & \\
\midrule
SD1 & Inter-spliced tiles & $1600$ & & \checkmark \\
\midrule
SD2 & Inter-spliced tiles & $3500$ & & & \checkmark \\
& Intra-spliced tiles & $3500$ & & & \checkmark \\
\bottomrule
\end{tabular}}
\end{table}

\noindent \textbf{Fingerprint Extraction Dataset (FED). }For creating this dataset, we took only tiles from the first 10 \gls{sar} products we downloaded. More specifically, we took the $50\%$ of tiles from each \gls{grd} product, reserving the remaining $50\%$ for creating the \gls{set1}. In this way we assured that the training of the \unet{} happened on samples never seen during training by the fingerprint extractor. 
In the end, we created a dataset of approximately $4000$ pristine tiles for training the fingerprint extractor. 

\noindent \textbf{Spliced Dataset 1 (SD1). }
The \gls{set1} is a dataset of splicing attacks created for training the \unet{}. 
The tiles 
have been taken from the first 10 \gls{grd} products downloaded for our experiments. More specifically, we took the $50\%$ of tiles not used for
training the fingerprint extractor. 

The splicing attacks have been realized in four different scenarios:
\begin{enumerate}
    \item with the donor tile $\mathbf{T}_D$ having undergone no editing;
    \item with the donor tile $\mathbf{T}_D$ having undergone a  rotation with angle chosen randomly;
    \item with the donor tile $\mathbf{T}_D$ having undergone resizing;
    \item with the donor tile $\mathbf{T}_D$ having undergone both rotation and resizing.
\end{enumerate}
For all four scenarios, we always considered the case where the donor tile $\mathbf{T}_D$ and the target tile $\mathbf{T}_T$ come from different products. 
We generated spliced tiles $\mathbf{T}_S$ with a spliced region $\mathcal{S}$ contained inside a $128\times 128$ or $256\times 256$ pixel area.
More specifically, we proceeded as follows:
\begin{enumerate}
    \item we applied a selected editing operation to $\mathbf{T}_D$;
    \item 
    we randomly cropped a pixel region from $\mathbf{T}_D$, imposing it to have a maximum resolution of either $128\times 128$ or $256\times 256$ pixels;
    \item we selected a random position in the target tile $\mathbf{T}_D$ and pasted the spliced region $\mathcal{S}$ on it.
\end{enumerate}
We considered different combination of parameters for the editing, resulting in a final number of $1600$ samples. For clarity's sake, Table~\ref{tab:spliced_datasets} reports all the considered editing operations with their parameters.
\begin{table*}[t]
\centering
\caption{Parameters used for the editing operated in the \gls{set1} and \gls{set2} and total number of samples. Parameters for noise-based editing are reported referring to samples with values between 0 and 1.}
\label{tab:spliced_datasets}
\resizebox{0.85\textwidth}{!}{
\begin{tabular}{ccccccc}
\toprule
\multicolumn{1}{c}{\textbf{Set}} &
  \textbf{Editing operation} &
   &
  \textbf{Editing parameters} &
  \multicolumn{1}{l}{\textbf{Total}} &
  \textbf{Total \# in set} \\ \midrule
SD1 & \textit{No editing}                  &  &                                                               & 200 & \multicolumn{1}{c}{$1600$} \\
    & \textit{Random rotation}             &  & Angle $\sim \mathcal{U}(\ang{-45}, \ang{45})$                     & $200$ &                          \\
    & \textit{Resize 1}                    &  & Factor $= 1.5$                                                & $200$ &                          \\
    & \textit{Resize 2}                    &  & Factor $= 2$                                                  & $200$ &                          \\
    & \textit{Resize 3}                    &  & Factor $= 2.5$                                                & $200$ &                          \\
    & \textit{Random rotation \& resize 1} &  & Angle $\sim \mathcal{U}(\ang{-45}, \ang{45})$; Factor $= 1.5$ & $200$ &                          \\
    & \textit{Random rotation \& resize 2} &  & Angle $\sim \mathcal{U}(\ang{-45}, \ang{45})$; Factor $= 2$   & $200$ &                          \\
    & \textit{Random rotation \& resize 3} &  & Angle $\sim \mathcal{U}(\ang{-45}, \ang{45})$; Factor $= 2.5$ & $200$ &                          \\ \midrule
SD2 & \textit{No editing}                  &  &                                                               & $1000$ & \multicolumn{1}{c}{$7000$} \\
    & \textit{Additive Gaussian noise}     &  & Mean $= 0$ ; Var $\sim \mathcal{U}(0, 0.1)$                     & $1000$ &                          \\
    & \textit{Additive Laplacian noise}    &  & Mean $= 0$; Var $\sim \mathcal{U}(0, 0.1)$                      & $1000$ &                          \\
    & \textit{Average blur}                &  & Kernel dim $=10\times 10$                                     & $1000$ &                          \\
    & \textit{Median blur}                 &  & Kernel dim $=5\times 5$                                       & $1000$ &                          \\
&  \textit{Random rotation \& resize} &
   &  Angle $\sim \mathcal{U}(\ang{-45}, \ang{45})$; Factor $\sim \mathcal{U}(1, 1.5)$ & $1000$ & \\
& \textit{Speckle-like noise}          &  & Mean $= 0$; Var $\sim \mathcal{U}(0, 0.3)$ & $1000$ & \\ \bottomrule
\end{tabular}}
\end{table*}

\noindent \textbf{Spliced Dataset 2 (SD2). }  The \gls{set2} is a second dataset of splicing attacks designed to test the performance of the complete pipeline. The composition of the \gls{set2} has been executed starting from the tiles of the last 10 \gls{grd} products at our disposal. These products have been reserved to this task to avoid any possible overlap between the data used for training the data-driven components of our pipeline, and the data used for testing them. 

For the generation of the \gls{set2}, we wanted a more challenging dataset with respect to the \gls{set1}. 
To do so, we considered both the cases where the donor tile $\mathbf{T}_D$ and the target tile $\mathbf{T}_T$ come from different or the same products. We define the first scenario as inter-splicing and the latter one as intra-splicing. Moreover, we extended the number of editing operations applied on  $\mathbf{T}_D$
using processing never seen by the \unet{}. 
For this last aspect, we tried to simulate an attacker perspective and considered operations that could make the tampering more plausible in the \gls{sar} imaging context. 

We used noise addition with two different distributions (Gaussian and Laplacian), two typologies of blurring (average, median), a similarity transformation comprehending rotation and scaling, a speckle-like multiplicative noise degradation and, finally, we considered also the case where no editing is applied to $\mathbf{T}_D$.
The parameters used for executing the editing
are all reported in Table~\ref{tab:spliced_datasets}.

We also varied the dimensions of the spliced region $\mathcal{S}$. Starting from the previous maximum pixel resolutions of $128\times 128$ and $256\times 256$ pixels, we included the intermediate areas of $160\times 160$, $192\times 192$ and $224\times 224$. 
In the end, $7000$ tiles compose the \gls{set2}, 
$3500$ realized in the inter-splicing scenario, and $3500$ realized in the intra-splicing one. These numbers account for $100$ spliced tiles per area and operation, multiplied by $2$ accounting for the inter and intra-splicing modalities.

\subsection{Training}
\label{subsec:training}

Here we briefly illustrate the training procedures followed for the data-driven components of our pipeline, i.e., the fingerprint extractor and the \unet{} mask estimator.

\subsubsection{Fingerprint Extraction}
\label{subsubsec:fingerprint_training}
The training set was constituted by the pristine tiles coming from the \gls{fed} dataset. 
For the fingerprint extractor, we relied on the mini-batch boost procedure originally employed by the authors in \cite{Cozzolino2020}. 
However, due to the limited amount of products available with respect to the original forensic task, we exploited more patches for the construction of the mini-batches. Specifically, in each batch we accounted for $4$ \gls{grd} products at a time, inserting $10$ tiles per product. From each tile, we randomly extracted $6$ patches of $48\times 48$ pixels, ending up with $240$ patches per mini-batch. 

In executing the training, we investigated three scenarios, leading to three different fingerprint extractors: 
\begin{itemize}
    \item \gls{be}: this extractor corresponds to training the fingerprint extractor proposed by \cite{Cozzolino2020} off-the-shelf on amplitude \gls{sar} images. Specifically, we trained the DnCNN 
    without relaxing the constraint 
    on the position of the patches to compute the \gls{dbl} loss (see Section~\ref{subsec:fingerprint} for details). 
    This extractor served as a baseline to evaluate the goodness of our proposed fingerprint extraction method.
    We randomly selected $5$ products for training and $5$ for validation, corresponding to $2000$ tiles for training and $2000$ for validation.
    \item \gls{sae}: in this scenario, we relaxed the patch position constraint 
    following the motivations reported in Section~\ref{subsec:fingerprint}. This translated into assigning a positive label in the \gls{dbl} loss to every patch pair coming from tiles of the same product, regardless of the position from which the patches have been extracted. As in \gls{be},
    we randomly picked
    $5$ products for training and $5$ for validation, ending up with $2000$ tiles for training and $2000$ for validation.
    \item \gls{asae}: for this extractor, 
    we relaxed the patch position constraint as done for the \gls{sae}, but we also applied data augmentation. 
    More specifically, we
    resized all the tiles using a $1.5$ scaling factor, and then randomly cropped them to $1024\times 1024$ pixels. As previously explained in Section~\ref{subsec:fingerprint},
    we considered all resized tiles as coming from separate \gls{grd} products.
    We exploited $20$ pristine products, i.e., $10$ original products and $10$ new products corresponding to their resized versions, randomly using $10$ for training (corresponding to $4000$ tiles), and $10$ (other $4000$ tiles) for validation. 
\end{itemize}
All the extractors have been trained for $500$ maximum epochs, with each epoch consisting of $128$ batch iterations, using Adam optimizer \cite{Kingma2014} with a learning rate of $10^{-4}$. 
We stopped the training if the validation loss did not improve for $30$ consecutive epochs. Then, we kept the model showing the best validation loss.

\subsubsection{U-Net Mask Estimator}
\label{subsec:unet_training}

Since the goal of the mask estimation task is highlighting potential forged areas in the fingerprint extracted from the query tile, the \unet{} training dataset must consist of fingerprints. 
Therefore, we extracted the fingerprints of the samples in the \gls{set1} by exploiting the three extractors listed in Section~\ref{subsubsec:fingerprint_training}.
Then, we trained a \unet{} on each set of fingerprints, creating a separate pipeline for each extractor.

We relied on the \unet{} model reported in \cite{Yakubovskiy:2019}, which allows to use various \glspl{cnn} as backbones for the encoder-decoder structure. 
Our choice fell on the EfficientNetB0 model \cite{pmlr-v97-tan19a}, a network of the EfficientNet family which proved extremely handy and recently found a discrete success both in multimedia forensics \cite{Bonettini2021, Mandelli2020, Yousfi2021} and in overhead imagery analysis \cite{Cannas2021}. We considered an EfficientNetB0 as encoder, and another one as decoder.

We randomly split the fingerprints extracted from the \gls{set1} dataset into $50\%$ for training and $50\%$ for validation.
We trained the networks for $500$ epochs, using as loss function the one described in Section~\ref{subsec:mask_estimation} and resorting to Adam optimization with a learning rate of $10^{-4}$.
We reduced the learning rate by a $0.1$ factor on plateau of the validation loss for $10$ consecutive epochs, and early-stopped the training if the validation loss did not improve for $30$ consecutive epochs. 
We kept the best validation model for all the networks trained with the three different fingerprint extractors.

\subsection{Tampering mask estimation parameters}
In order for our unsupervised methods (i.e., K-means and \gls{gmm}) to be successfully deployed, the dimension of the extracted patches $\mathbf{P}_n$, as well as the number $C$ of clusters in which the fingerprint $\mathbf{F}$ is partitioned, are crucial aspects. A too big resolution of each $\mathbf{P}_n$ or a small number of clusters $C$ might lead to an under partition of the fingerprint, which is generally way less preferable with respect to an over segmentation. 

For this reason, we spent a preliminary part of our work in determining the right amount of clusters and the right patches resolution, finding a good trade-off in dividing the fingerprint $\mathbf{F}$ into non-overlapping patches $8\times 8$ pixels wide,
and using $7$ clusters.
Also the \unet{} needs to have a correct threshold $\tau$ applied to the probability mask $\mathbf{\hat{M}}^u$. In this case, we found an optimal value with $\tau=0.5$.

\subsection{Evaluation metrics}
\begin{figure}[t]
\caption{IOU localization examples. The blue square represents the ground-truth area, while the orange square shows the predicted area. }
\label{fig:IOU_examples}
\centering
\subfloat[$\textrm{IOU} \sim 0.2$.]{\includegraphics[width=0.2\columnwidth]{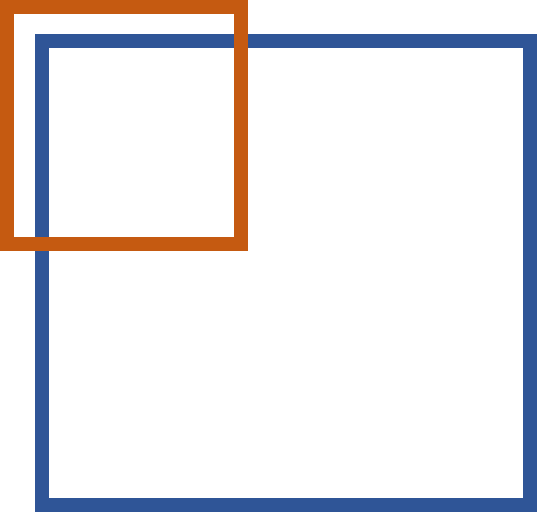}
\label{fig:poor_IOU}}
\hfil
\subfloat[$\textrm{IOU} \sim0.7$.]{\includegraphics[width=0.2\columnwidth]{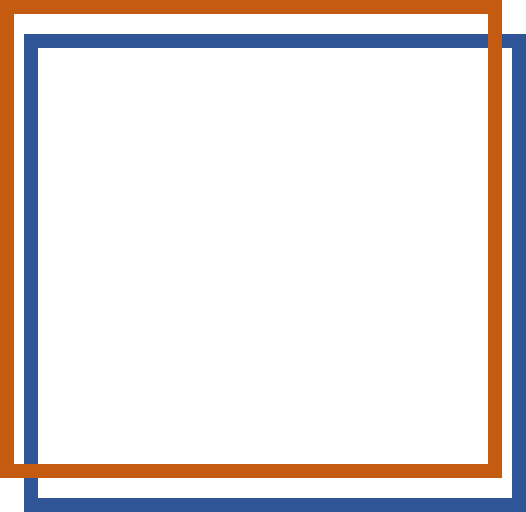}
\label{fig:good_IOU}}
\hfil
\subfloat[$\textrm{IOU} \sim 0.9$.]{\includegraphics[width=0.2\columnwidth]{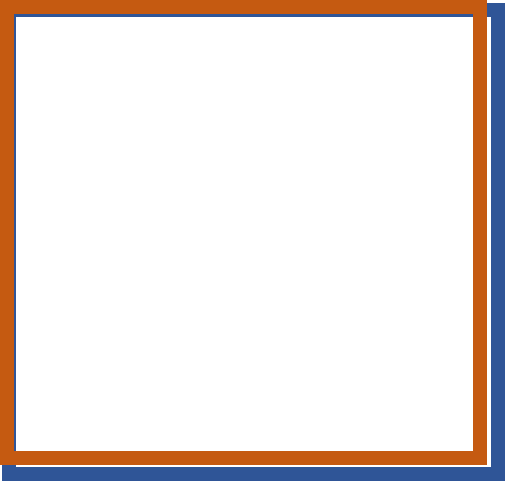}
\label{fig:excellent_IOU}}
\end{figure}
For evaluating our performances in correctly estimating the tampering mask, we relied on two metrics: the balanced accuracy and the Jaccard index or \gls{iou}.
Given an estimated tampering mask $\mathbf{\hat{M}}$, we can divide its pixels based on the correctness of the tampering localization. Specifically, we can assign them to four different categories:
\begin{itemize}
    \item \gls{tp}: spliced pixels classified as spliced;
    \item \gls{fp}: pristine pixels classified as spliced;
    \item \gls{tn}: pristine pixels classified as pristine;
    \item \gls{fn}: spliced pixels classified as pristine.
\end{itemize}
The balanced accuracy is defined as:
\begin{equation}
    \label{balacc}
    \textrm{BA} = \frac{1}{2} \Bigl(\frac{\textrm{\gls{tp}}}{\textrm{\gls{tp}}+\textrm{\gls{fn}}} + \frac{\textrm{\gls{tn}}}{\textrm{\gls{tn}}+\textrm{\gls{fp}}}\Bigr).
\end{equation}
This quantity 
measures how well our pipeline performed in correctly assigning each pixel in the estimated tampering mask $\mathbf{\hat{M}}$, taking into account the disproportion between pristine and spliced pixels. The higher the balanced accuracy, the better the splicing localization.

The \gls{iou} is defined as:
\begin{equation}
    \textrm{\gls{iou}} = \frac{\textrm{\gls{tp}}}{\textrm{\gls{tp}}+\textrm{\gls{fp}}+\textrm{\gls{fn}}}.
\end{equation}
This measure is popular in computer vision for object detection tasks, where it is used to quantify how well a predicted bounding-box for an object overlaps with the actual object's position. For our task, this translates in the \gls{iou} accurately quantifying how well the area localized in the estimated tampering mask $\mathbf{\hat{M}}$ overlaps with the one indicated in the original mask $\mathbf{M}$. Fig.~\ref{fig:IOU_examples} shows some examples. Values close to $1$ are better, but an \gls{iou} equal or greater than $0.5$ is good too.

%% file: sec_results.tex
\section{Results}
\label{sec:results}

\begin{table*}[t]
\centering
\caption{Average localization results per operation and per binary mask estimation method in the inter-splicing scenario. Best results in bold. }
\label{tab:localization_results_all_methods_copy_paste}
\begin{tabular}{@{}ccccccccc@{}}
\toprule
\textbf{Operation} & \textbf{Fingerprint extractor} & \multicolumn{3}{c}{\textbf{IOU per method}} & \multicolumn{3}{c}{\textbf{BA per method}}\\ 
& & \textit{\gls{gmm}} & \textit{K-means} & \textit{\unet{}} & \textit{\gls{gmm}} & \textit{K-means} & \textit{\unet{}} \\ 
\midrule
\textit{No editing} & \gls{be} (baseline) & $0.232$ & $0.156$ & $0.185$ & $0.643$ & $0.605$ & $0.640$ \\
& \gls{sae} & $0.163$ & $0.036$ & $0.241$ & $0.595$ & $0.527$ & $0.691$ \\
& \gls{asae} & $0.163$ & $0.044$ & $\mathbf{0.258}$ & $0.594$ & $0.530$ & $\mathbf{0.697}$ \\ 
\midrule
\textit{Additive Gaussian noise} & \gls{be} (baseline) & $0.462$ & $0.243$ & $0.297$ & $0.787$ & $0.687$ & $0.772$ \\
& \gls{sae} & $0.459$ & $0.248$ & $0.200$ & $0.777$ & $0.747$ & $0.664$ \\
& \gls{asae} & $\mathbf{0.669}$ & $0.544$ & $0.433$ & $0.867$ & $0.853$ & $\mathbf{0.913}$ \\ 
\midrule
\textit{Additive Laplacian noise} & \gls{be} (baseline) & $0.481$ & $0.264$ & $0.300$ & $0.795$ & $0.687$ & $0.772$ \\
& \gls{sae} & $0.497$ & $0.266$ & $0.174$ & $0.804$ & $0.761$ & $0.629$ \\
& \gls{asae} & $\mathbf{0.722}$ & $0.596$ & $0.451$ & $0.895$ & $0.891$ & $\mathbf{0.914}$ \\ 
\midrule
\textit{Average blur} & \gls{be} (baseline) & $0.748$ & $0.143$ & $0.214$ & $0.899$ & $0.588$ & $0.844$ \\
& \gls{sae} & $0.603$ & $0.539$ & $0.645$ & $0.889$ & $0.800$ & $0.939$ \\
& \gls{asae} & $\mathbf{0.859}$ & $0.704$ & $0.697$ & $0.940$ & $0.860$ & $\mathbf{0.972}$ \\ 
\midrule
\textit{Median blur} & \gls{be} (baseline) & $0.517$ & $0.206$ & $0.368$ & $0.803$ & $0.649$ & $0.885$ \\
& \gls{sae} & $0.685$ & $0.513$ & $0.651$ & $0.876$ & $0.783$ & $0.941$ \\
& \gls{asae} & $\mathbf{0.811}$ & $0.627$ & $0.694$ & $0.918$ & $0.816$ & $\mathbf{0.972}$ \\ 
\midrule
\textit{Random rotation \& resize} & \gls{be} (baseline) & $0.406$ & $0.228$ & $0.368$ & $0.734$ & $0.674$ & $0.882$ \\
& \gls{sae} & $0.417$ & $0.174$ & $0.625$ & $0.745$ & $0.590$ & $0.952$ \\
& \gls{asae} & $0.593$ & $0.379$ & $\mathbf{0.681}$ & $0.823$ & $0.714$ & $\mathbf{0.972}$ \\ 
\midrule
\textit{Speckle-like noise} & \gls{be} (baseline) & $0.503$ & $0.259$ & $0.306$ & $0.809$ & $0.706$ & $0.798$ \\
& \gls{sae} & $0.428$ & $0.219$ & $0.154$ & $0.778$ & $0.742$ & $0.617$ \\
& \gls{asae} & $\mathbf{0.721}$ & $0.568$ & $0.495$ & $0.900$ & $0.895$ & $\mathbf{0.942}$ \\ 
\bottomrule
\end{tabular}
\end{table*}
\begin{table*}[h!]
\centering
\caption{Average localization results per operation and per binary mask estimation method in the intra-splicing scenario. Best results in bold. }
\label{tab:localization_results_all_methods_copy_move}
\begin{tabular}{@{}ccccccccc@{}}
\toprule
\textbf{Operation} & \textbf{Fingerprint extractor} & \multicolumn{3}{c}{\textbf{IOU per method}} & \multicolumn{3}{c}{\textbf{BA per method}}\\ 
& & \textit{\gls{gmm}} & \textit{K-means} & \textit{\unet{}} & \textit{\gls{gmm}} & \textit{K-means} & \textit{\unet{}} \\ 
\midrule
\textit{No editing} & \gls{be} (baseline) & $0.024$ & $0.021$ & $0.027$ & $0.502$ & $0.500$ & $0.510$ \\
& \gls{sae} & $0.025$ & $0.024$ & $0.022$ & $0.501$ & $0.502$ & $0.512$ \\
& \gls{asae} & $\mathbf{0.028}$ & $0.026$ & $0.015$ & $0.502$ & $0.501$ & $\mathbf{0.516}$ \\ 

\midrule
\textit{Additive Gaussian noise} & \gls{be} (baseline) & $0.300$ & $0.112$ & $0.212$ & $0.699$ & $0.609$ & $0.701$ \\
& \gls{sae} & $0.420$ & $0.191$ & $0.093$ & $0.756$ & $0.706$ & $0.557$ \\
& \gls{asae} & $\mathbf{0.602}$ & $0.482$ & $0.390$ & $0.834$ & $0.837$ & $\mathbf{0.854}$ \\ 
\midrule
\textit{Additive Laplacian noise} & \gls{be} (baseline) & $0.364$ & $0.186$ & $0.234$ & $0.738$ & $0.662$ & $0.707$ \\
& \gls{sae} & $0.408$ & $0.195$ & $0.106$ & $0.758$ & $0.705$ & $0.562$ \\
& \gls{asae} & $\mathbf{0.663}$ & $0.532$ & $0.426$ & $0.869$ & $0.864$ & $\mathbf{0.887}$ \\ 
\midrule
\textit{Average blur} & \gls{be} (baseline) & $0.700$ & $0.082$ & $0.274$ & $0.881$ & $0.544$ & $0.894$ \\
& \gls{sae} & $0.478$ & $0.616$ & $0.645$ & $0.819$ & $0.825$ & $0.943$ \\
& \gls{asae} & $\mathbf{0.791}$ & $0.764$ & $0.709$ & $0.906$ & $0.882$ & $\mathbf{0.953}$ \\ 
\midrule
\textit{Median blur} & \gls{be} (baseline) & $0.584$ & $0.310$ & $0.468$ & $0.838$ & $0.714$ & $0.926$ \\
& \gls{sae} & $0.694$ & $0.597$ & $0.674$ & $0.859$ & $0.811$ & $\mathbf{0.957}$ \\
& \gls{asae} & $\mathbf{0.787}$ & $0.684$ & $0.717$ & $0.899$ & $0.842$ & $0.956$ \\ 
\midrule
\textit{Random rotation \& resize} & \gls{be} (baseline) & $0.344$ & $0.323$ & $0.418$ & $0.711$ & $0.722$ & $0.925$ \\
& \gls{sae} & $0.328$ & $0.265$ & $0.609$ & $0.712$ & $0.646$ & $0.949$ \\
& \gls{asae} & $0.543$ & $0.497$ & $\mathbf{0.671}$ & $0.787$ & $0.782$ & $\mathbf{0.953}$ \\ 
\midrule
\textit{Speckle-like noise} & \gls{be} (baseline) & $0.352$ & $0.098$ & $0.273$ & $0.728$ & $0.623$ & $0.776$ \\
& \gls{sae} & $0.366$ & $0.164$ & $0.035$ & $0.747$ & $0.711$ & $0.519$ \\
& \gls{asae} & $\mathbf{0.615}$ & $0.501$ & $0.486$ & $0.857$ & $0.869$ & $\mathbf{0.916}$ \\ 
\bottomrule
\end{tabular}
\end{table*}
In this section, we report the results of our experimental campaign. 
In particular, we describe the achieved results for the splicing localization task and compare our performances with state-of-the-art. 


To have a fair comparison among the different fingerprint extraction and mask estimation methods,
we resort to dataset \gls{set2} for evaluating all the results. 
It is worth noticing that \gls{set2} comprehends acquisitions never seen by any of the data-driven blocks of our pipeline, together with new unseen editing operations.
We show our results by 
considering separated the two different scenarios of donor and target tiles coming from the same (i.e., intra-splicing) or different (i.e., inter-splicing) acquired products. 

\begin{figure}[t]
\centering
\subfloat[Spliced tile $\mathbf{T}_S$.]{\includegraphics[width=0.3\columnwidth]{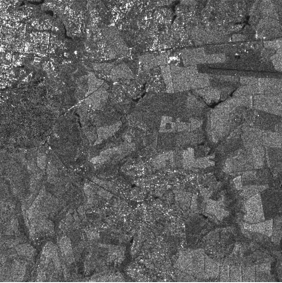}%
\label{fig:pipeline_example_0:for_0}}
\hfil
\subfloat[Tampering mask $\mathbf{M}$.]{\includegraphics[width=0.3\columnwidth]{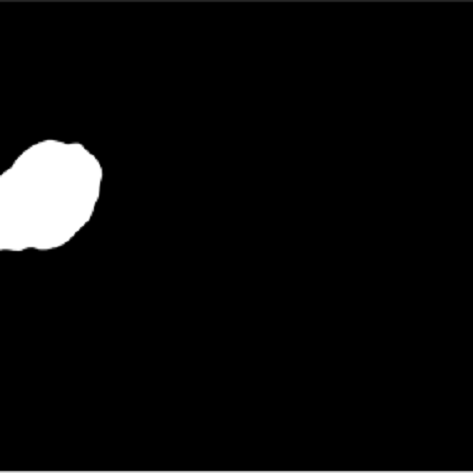}%
\label{fig:pipeline_example_0:mask_0}}
\hfil
\subfloat[Fingerprint $\mathbf{F}$.]{\includegraphics[width=0.3\columnwidth]{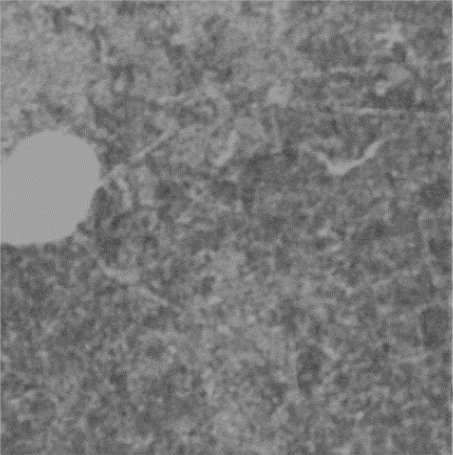}%
\label{fig:pipeline_example_0:fp_0}}
\\
\subfloat[\gls{gmm} estimated tampering mask $\hat{\mathbf{M}}$.]{\includegraphics[width=0.3\columnwidth]{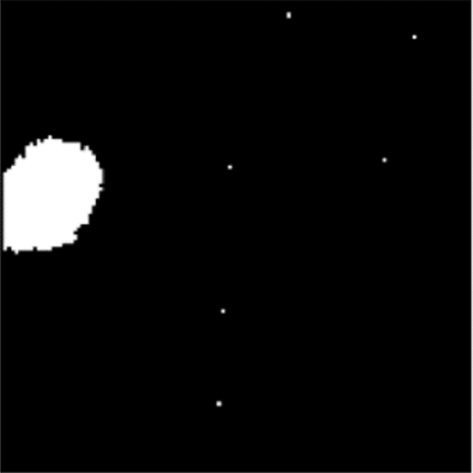}%
\label{fig:pipeline_example_0:gmm_0}}
\hfil
\subfloat[K-means estimated tampering mask $\hat{\mathbf{M}}$.]{\includegraphics[width=0.3\columnwidth]{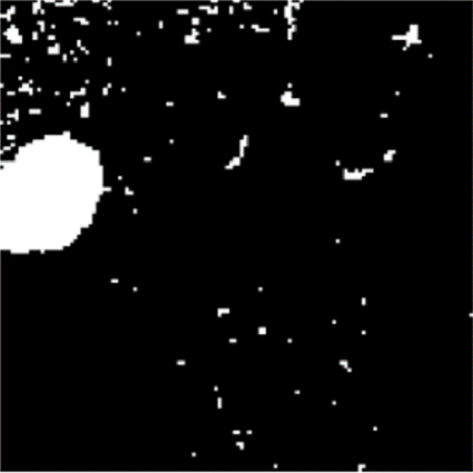}%
\label{fig:pipeline_example_0:kmeans_0}}
\hfil
\subfloat[\unet{} estimated tampering mask $\hat{\mathbf{M}}$.]{\includegraphics[width=0.3\columnwidth]{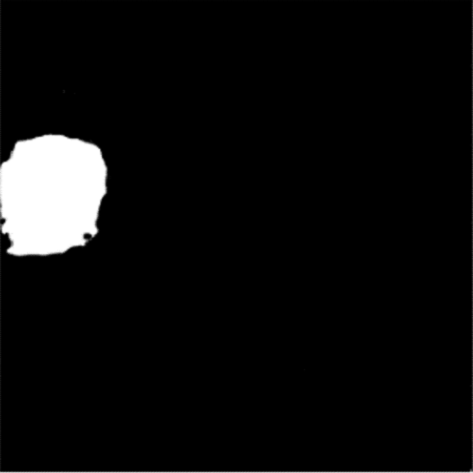}%
\label{fig:pipeline_example_0:unet_0}}
\caption{Example of spliced tile with resizing and gaussian noise addition applied to the donor tile $\mathbf{T}_D$. The pipeline is based on the \gls{asae}.}
\label{fig:pipeline_example_0}
\end{figure}

\begin{figure}[t]
\centering
\subfloat[Spliced tile $\mathbf{T}_S$.]{\includegraphics[width=0.3\columnwidth]{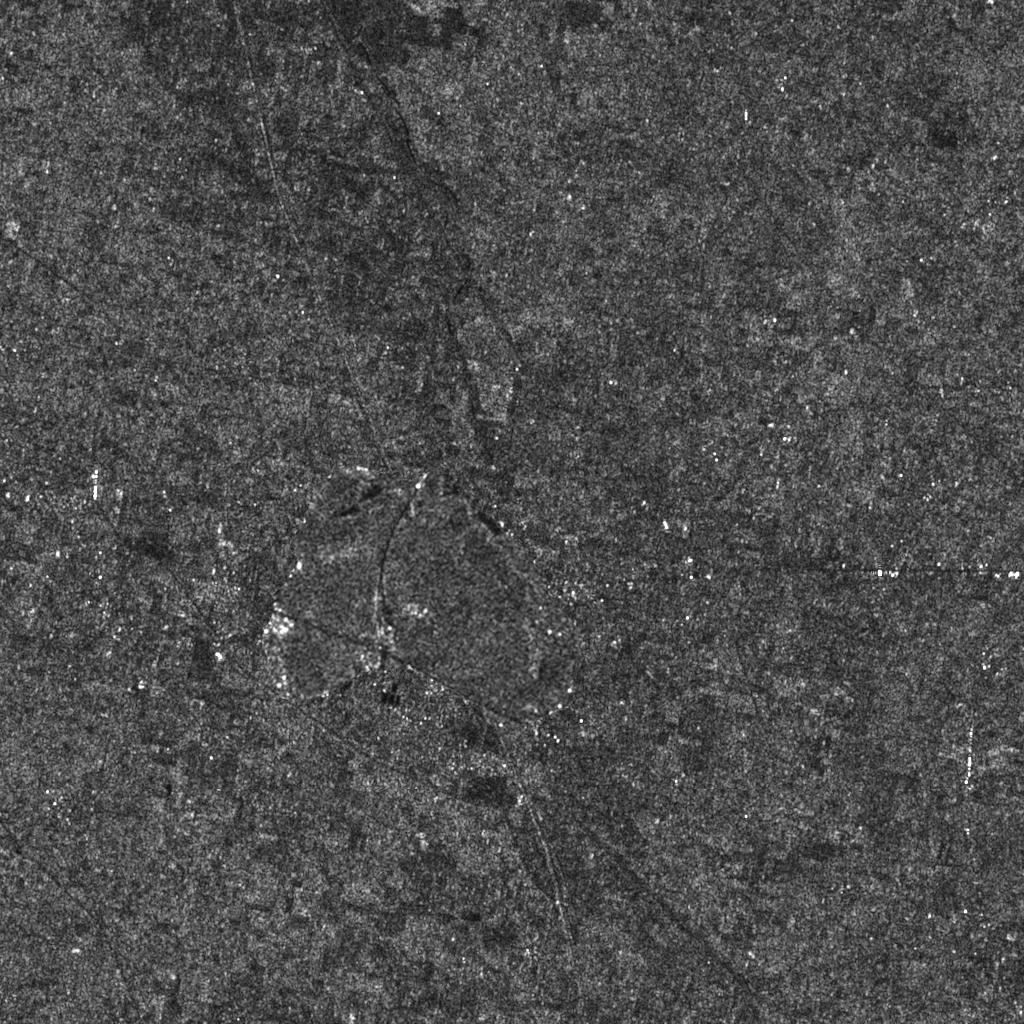}%
\label{fig:pipeline_example_1:for_0}}
\hfil
\subfloat[Tampering mask $\mathbf{M}$.]{\includegraphics[width=0.3\columnwidth]{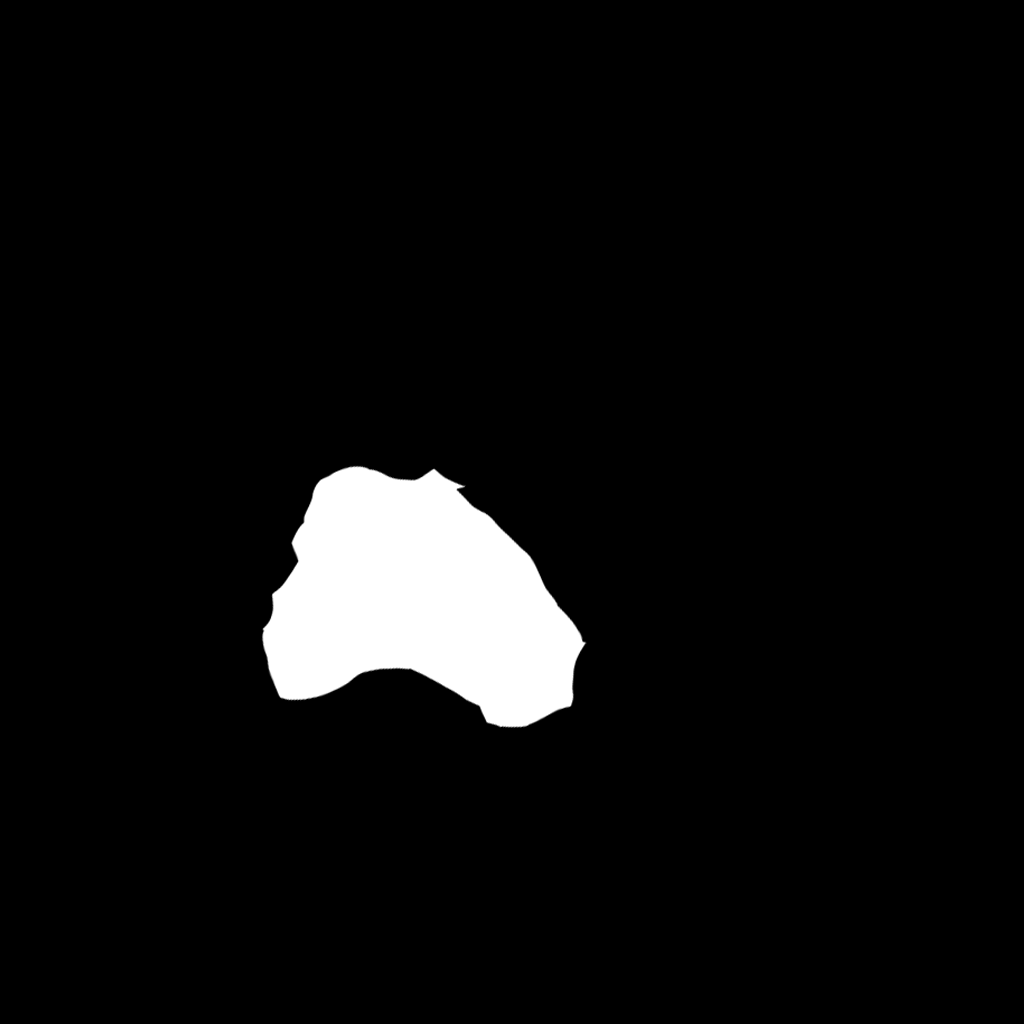}%
\label{fig:pipeline_example_1:mask_0}}
\hfil
\subfloat[Fingerprint $\mathbf{F}$.]{\includegraphics[width=0.3\columnwidth]{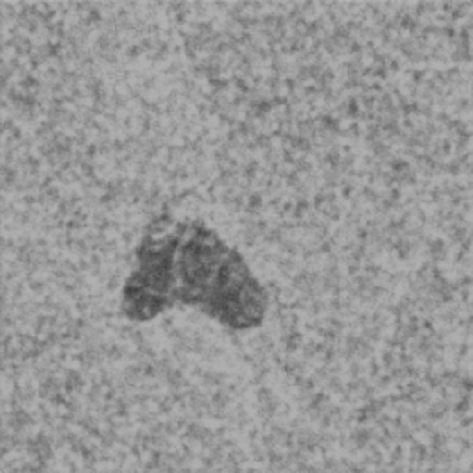}%
\label{fig:pipeline_example_1:fp_0}}
\\
\subfloat[\gls{gmm} estimated tampering mask $\hat{\mathbf{M}}$.]{\includegraphics[width=0.3\columnwidth]{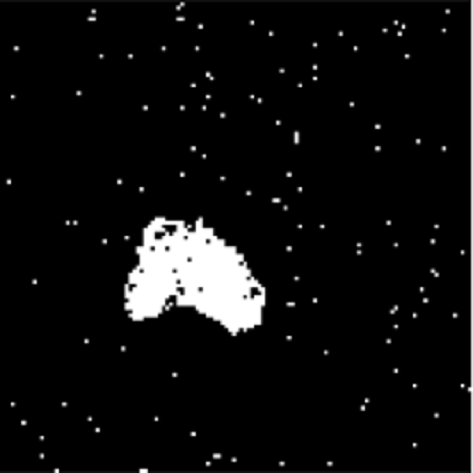}%
\label{fig:pipeline_example_1:gmm_0}}
\hfil
\subfloat[K-means estimated tampering mask $\hat{\mathbf{M}}$.]{\includegraphics[width=0.3\columnwidth]{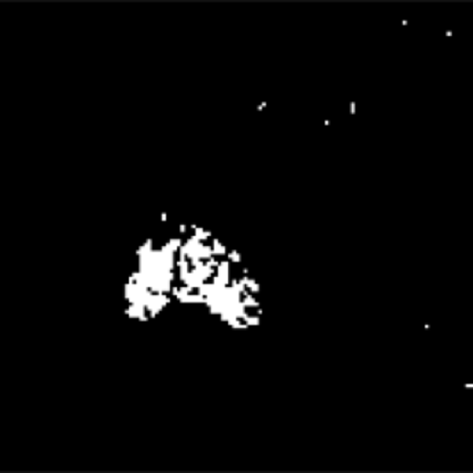}%
\label{fig:pipeline_example_1:kmeans_0}}
\hfil
\subfloat[\unet{} estimated tampering mask $\hat{\mathbf{M}}$.]{\includegraphics[width=0.3\columnwidth]{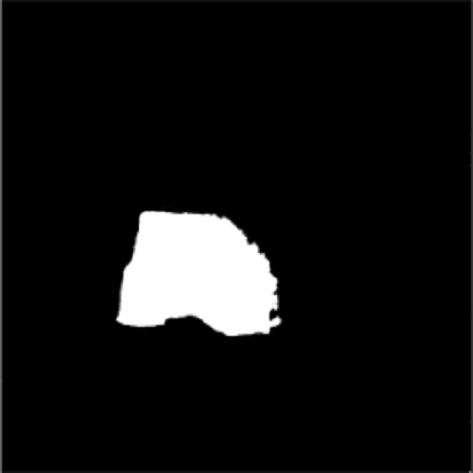}%
\label{fig:pipeline_example_1:unet_0}}
\caption{Example of spliced tile. An urban area is covered with resizing applied to the donor tile $\mathbf{T}_D$. The pipeline used is based on the \gls{asae}.}
\label{fig:pipeline_example_1}
\end{figure}

Tables \ref{tab:localization_results_all_methods_copy_paste} and \ref{tab:localization_results_all_methods_copy_move} report the localization results by combining the two proposed fingerprint extractors (i.e., \gls{sae} and \gls{asae}) and the baseline fingerprint extractor (i.e., \gls{be}). 
Specifically, Table~\ref{tab:localization_results_all_methods_copy_paste} depicts the results achieved in the inter-splicing scenario. On the contrary, Table~\ref{tab:localization_results_all_methods_copy_move} shows results for the intra-splicing scenario. 
Finally, Figures~\ref{fig:pipeline_example_0} and \ref{fig:pipeline_example_1} report some examples of splicing attacks together with all the artifacts generated by our proposed pipeline. In the following we report the major findings from these results.

\subsection{Fingerprint extractors comparison}
The best fingerprint extractor is always the \gls{asae}, with the mask estimation methods \gls{gmm} and \unet{} alternating in providing the best performances. 
Moreover, the \gls{sae} showed on average better results than the baseline extractor, on $4$ editing operations out of $7$ in the inter-splicing scenario and on $6$ operations out of $7$ in the intra-splicing scenario.

Notice that, while the 
relaxation of the patch position constraint proposed in Section~\ref{subsec:fingerprint} (i.e., the \gls{sae} configuration) provided us better average metrics with respect to the baseline, the additional insertion of a simple data augmentation like resizing (i.e., the \gls{asae} configuration) gave us an even greater performance boost. From this point of view, it is worth mentioning that the \gls{asae}-based detectors showed better results also on editing operations which are not strictly related to resizing. This is true for the noise addition and blurring operations, for instance.
\color{black}

\subsection{``No editing'': inter-splicing versus intra-splicing. }
On the ``No editing'' operation, all the pipelines presented the worst results. 
Moreover, we observed significant differences in performances depending on the donor and target tiles coming from different (i.e., inter-splicing) or the same (i.e., intra-splicing) acquired products. 
In the first scenario, with best \gls{iou} and balanced accuracy of $0.25$ and $0.69$ respectively, performances were still fairly good.
In the second scenario, 
the evaluation metrics depicted instead an almost random decision for the estimation of the tampering mask.

This different behavior was somehow expected. 
As explained in Section~\ref{sec:method}, our proposed pipeline has been designed with the objective of capturing inconsistencies related to the generation process of amplitude \gls{sar} products. 
The fingerprint extraction has been trained to provide a globally incoherent fingerprint 
only if the spliced region 
and its surrounding areas have undergone different processing. 
In intra-splicing attacks, when no editing is applied, splicing inconsistencies are absent as $\mathbf{T}_D$ and $\mathbf{T}_T$ come from the same GRD product. 
For this reason, we expected our detectors not being able to localize such 
attacks. 
\subsection{Generalization on editing operations}
With the only exception of the ``No editing'' scenario, in Table~\ref{tab:localization_results_all_methods_copy_paste}, the results achieved on each editing operation always depict an \gls{iou} greater than $0.66$ and a balanced accuracy exceeding $0.91$. 
In Table~\ref{tab:localization_results_all_methods_copy_move}, we exceeded $0.60$ and $0.85$ for \gls{iou} and balanced accuracy, respectively.
While intra-splicing results are lower than inter-splicing, we expected such a behavior for the reasons reported above: spliced tiles in intra-splicing modality do not present inconsistencies in the forensic traces related to the pipeline that generated their original products, making them a more difficult asset to analyze. 
However, the overall good performances also in the intra-splicing scenario suggest the proposed pipeline was useful in finding inconsistencies associated to general editing operations executed on $\mathcal{S}$.

It is interesting to notice that all the methods performed consistently across the different types of editing considered (i.e., noise addition, blurring, rotation, resizing, noise multiplication). In particular, the results achieved by the \unet, which showed the best balanced accuracy for noise-based attacks, were quite surprising, considering that the \unet{} was trained only on resizing-based attacks. 
\begin{table*}[t]
\centering
\caption{Best localization results per operation executed on the \gls{set2} tiles in the inter-splicing scenario. 
The best tampering mask estimation method has been chosen from table \ref{tab:localization_results_all_methods_copy_paste}. 
Best results in bold. }
\label{tab:localization_results_best_method_copy_paste}
\begin{tabular}{@{}cccccccccc@{}}
\toprule
\textbf{Operation} & \multicolumn{4}{c}{\textbf{Best IOU per extractor}} & \multicolumn{4}{c}{\textbf{Best BA per extractor}}\\ 
& \textit{\gls{be} (Noiseprint)} & \textit{\gls{sae}} & \textit{\gls{asae}} & \textit{Splicebuster} & \textit{\gls{be} (Noiseprint)} & \textit{\gls{sae}} & \textit{\gls{asae}} & \textit{Splicebuster} \\
\midrule
\textit{No editing} & $0.232$ & $0.241$ & $\mathbf{0.258}$ & $0.194$ & $0.643$ & $0.691$ & $\mathbf{0.697}$ & $0.619$ \\ 
\midrule
\textit{Additive Gaussian noise} & $0.462$ & $0.459$ & $\mathbf{0.669}$ & $0.550$ & $0.787$ & $0.777$ & $\mathbf{0.913}$ & $0.799$ \\ 
\midrule
\textit{Additive Laplacian noise} & $0.481$ & $0.497$ & $\mathbf{0.722}$ & $0.552$ & $0.795$ & $0.804$ & $\mathbf{0.914}$ & $0.796$ \\ 
\midrule
\textit{Average blur} & $0.748$ & $0.645$ & $\mathbf{0.859}$ & $0.478$ & $0.899$ & $0.932$ & $\mathbf{0.972}$ & $0.808$ \\ 
\midrule
\textit{Median blur} & $0.517$ & $0.685$ & $\mathbf{0.811}$ & $0.508$ & $0.885$ & $0.941$ & $\mathbf{0.972}$ & $0.770$ \\ 
\midrule
\textit{Random rotation \& resize} & $0.406$ & $0.625$ & $\mathbf{0.681}$ & $0.551$ & $0.882$ & $0.952$ & $\mathbf{0.972}$ & $0.753$ \\ 
\midrule
\textit{Speckle-like noise} & $0.503$ & $0.428$ & $\mathbf{0.721}$ & $0.537$ & $0.809$ & $0.778$ & $\mathbf{0.942}$ & $0.781$ \\ 
\bottomrule
\end{tabular}
\end{table*}

\begin{table*}[t]
\centering
\caption{Best localization results per operation executed on the \gls{set2} tiles in the intra-splicing scenario. The best tampering mask estimation method has been chosen from table \ref{tab:localization_results_all_methods_copy_move}. Best results in bold.}
\label{tab:localization_results_best_method_copy_move}
\begin{tabular}{@{}cccccccccc@{}}
\toprule
\textbf{Operation} & \multicolumn{4}{c}{\textbf{Best IOU per extractor}} & \multicolumn{4}{c}{\textbf{Best BA per extractor}}\\ & \textit{\gls{be} (Noiseprint)} & \textit{\gls{sae}} & \textit{\gls{asae}} & \textit{Splicebuster} & \textit{\gls{be} (Noiseprint)} & \textit{\gls{sae}} & \textit{\gls{asae}} & \textit{Splicebuster} \\

\midrule
\textit{No editing} & $0.027$ & $0.025$ & $\mathbf{0.028}$ & $0.017$ & $0.510$ & $0.512$ & $\mathbf{0.516}$ & $0.481$ \\ 
\midrule
\textit{Additive Gaussian noise} & $0.300$ & $0.420$ & $\mathbf{0.602}$ & 0.575 & $0.701$ & $0.756$ & $\mathbf{0.854}$ & $0.814$ \\ 
\midrule
\textit{Additive Laplacian noise} & $0.364$ & $0.408$ & $\mathbf{0.663}$ & $0.599$ & $0.738$ & $0.758$ & $\mathbf{0.887}$ & 0.824 \\ 
\midrule
\textit{Average blur} & $0.700$ & $0.645$ & $\mathbf{0.791}$ & 0.524 & $0.894$ & $0.943$ & $\mathbf{0.953}$ & $0.786$ \\ 
\midrule
\textit{Median blur} & $0.584$ & $0.694$ & $\mathbf{0.787}$ & $0.661$ & $0.926$ & $\mathbf{0.957}$ & $0.956$ & $0.770$ \\ 
\midrule
\textit{Random rotation \& resize} & 0.418 & 0.609 & $\mathbf{0.671}$ & 0.559 & 0.925 & 0.949 & $\mathbf{0.953}$ & $0.807$ \\ 
\midrule
\textit{Speckle-like noise} & $0.352$ & $0.366$ & $\mathbf{0.615}$ & $0.606$ & $0.776$ & $0.747$ & $\mathbf{0.916}$ & $0.823$ \\ 
\bottomrule
\end{tabular}
\end{table*}

\subsection{Supervised versus unsupervised approaches}
Comparing the different mask estimation methods, we can notice that \gls{gmm} and \unet{} alternated in providing the best performances. 
The \unet{} represented the best method for all editing operations in terms of balanced accuracy, while the \gls{gmm} provided best results in terms of \gls{iou} on 5 operations out of 7. 
The K-means, while being the worst method of the three, showed nevertheless fairly good results. 
For instance, it was better than the \unet{} in 5 operations out of 7 in terms of \gls{iou} in both inter and intra-splicing scenarios.

While the \unet{}-based strategy might seem the most promising one on average, we must be aware that, as a supervised technique, it needed a preliminary stage of training. 
Since the training of \glspl{cnn} takes time and computational resources, unsupervised methods may be appealing as well, depending on the final needs and the resources at disposal.
As unsupervised methods, they are faster to deploy while still providing good performances.

\subsection{Comparison with state-of-the-art}
For what concerns the comparison with state-of-the-art,
Tables \ref{tab:localization_results_best_method_copy_paste} and \ref{tab:localization_results_best_method_copy_move} summarize the best localization results achieved by the two proposed fingerprint extractors for each editing operation, along with the results obtained by the Noiseprint method \cite{Cozzolino2020} and the Splicebuster method \cite{Cozzolino2015}. 
We selected these techniques as they are widely exploited as a baseline in the forensics literature.
Moreover, they proved to be robust to standard editing operations and do not require many adaptations to the domain of data under investigation.  

Since the Noiseprint produces real-valued heatmaps (without binarization), all the tampering masks 
have been created starting from the extracted fingerprint 
and following our proposed mask estimation methods (i.e., the K-means, \gls{gmm} and \unet{}). 
The achieved results exactly correspond to the \gls{be} results that we have previously shown. Splicebuster returns real-valued heatmaps as well, but
in this case we estimated binary tampering masks following
the methodology suggested by the same authors in \cite{Cozzolino2016}.


The state-of-the-art results 
always showed inferior performances than our best localization method.
Nonetheless, especially looking at the editing operations involving noise addition or multiplication (i.e., the Speckle-like noise), Splicebuster presented even better performances than methods based on \gls{be} and \gls{sae} extractors. Despite the clear differences between natural images and amplitude \gls{sar} products, from the nature of the signals they depict to the different processing 
that leads to their formation, such results seem to indicate that forensics tools based on generic footprints 
might reveal to be useful also in the \gls{sar} context. This is especially true if the attacker relies on common editing operations (e.g., resizing, blurring, etc.), where features like the high-pass frequency co-occurrences used by Splicebuster are robust enough to be an effective splicing localization tool.

%% file: sec_conclusions.tex
\section{Conclusions}\label{sec:conclusions}
In this paper, we analyzed the problem of splicing localization in amplitude \gls{sar} imagery. 
The forensic analysis of these objects is becoming of paramount relevance, as amplitude \gls{sar} products 
are relatively easy to handle and process, even with general editing software such as GIMP or Photoshop.

To the best of our knowledge, no solution has been proposed yet in the forensics literature tailored to this kind of signals.
As a matter of fact, amplitude \gls{sar} products present a completely different nature with respect to natural imagery, therefore are
posing new and different challenges in assessing their integrity. 

Inspired by a state-of-the-art method developed for natural images,
we proposed a new splicing localization technique specifically designed for amplitude \gls{sar} products. 
Our proposed method extracts a fingerprint localizing spliced regions in \gls{sar} tiles. Then, the fingerprint can be analyzed using three different methods, one supervised and two unsupervised, to generate a final tampering mask. This mask is a binary heatmap indicating whether pixels underwent splicing or not. 
We generated different datasets of spliced \gls{grd} tiles, 
to train and test the validity of our proposed method. We considered different kind of manipulations applied to the tiles, from noise-based attacks to blurring and resizing.

All proposed techniques showed encouraging results in the localization of splicing attacks, providing better performances when compared to state-of-the-art solutions developed for 
natural images.
The supervised approach 
reported the best numbers in terms of balanced accuracy, however needing a preliminary stage of training. 
The unsupervised approaches 
showed instead better performances in terms of the \gls{iou} metrics, and while they are less accurate in terms of balanced accuracy, they do not require a training phase.

These results proved the feasibility of the forensic analysis of amplitude \gls{sar} imagery, paving the way to further investigations on the development of methods tailored to this kind of signals.
Possible research themes regard the evaluation of the proposed pipeline over elaborated splicing attacks (e.g., \gls{gan}-generated inpainting) that should in principle be detected by our technique, a specific exploitation of traces related to the generation pipeline of \gls{sar} images, the use of physics-based clues linked to the scene represented in the data and, finally, the adaptation of splicing localization methods for electrical-optical imagery.